\documentclass[12pt,titlepage]{article}
\pdfoutput=1

\usepackage{authblk}
\usepackage{chemformula} 
\usepackage[T1]{fontenc} 
\usepackage{graphicx} 
\usepackage{color} 
\usepackage{ulem}
\usepackage{amssymb}
\usepackage{amsmath}
\usepackage{lipsum}
\usepackage{float}

\begin{document}

\title{Growth at high substrate coverage can decrease
the grain boundary roughness of 2D materials}

\author{Fabio D. A. Aar\~ao Reis}
\affil{Instituto de F\'{i}sica, Universidade Federal Fluminense,
Avenida Litor\^{a}nea s/n, 24210-340 Niter\'{o}i, RJ, Brazil\\
fdaar@protonmail.com}
\author{Bastien Marguet}
\affil{Institut Lumi\`ere Mati\`ere, UMR5306 Universit\'e Lyon 1 - CNRS,
Villeurbanne 69622, France\\
bastien.marguet@univ-lyon1.fr}
\author{Olivier Pierre-Louis}
\affil{Institut Lumi\`ere Mati\`ere, UMR5306 Universit\'e Lyon 1 - CNRS,
Villeurbanne 69622, France\\
olivier.pierre-louis@univ-lyon1.fr}

\maketitle

\begin{abstract}
Grain boundary roughness can affect
electronic and mechanical properties of two-dimensional materials. This roughness
depends crucially on the growth process by which the two-dimensional material is formed.
To investigate the key mechanisms that govern the roughness,
we have performed kinetic Monte Carlo simulations of a simple model 
that includes particle attachment, detachment, and diffusion.
We have studied the closure of the gap between two flakes during growth,
and the subsequent formation of the grain boundary (GB) 
for a broad range of model parameters.
The well known near-equilibrium (attachment-limited) and unstable (diffusion-limited) growth regimes are identified, but we also observe a third regime when the precursor flux is sufficiently 
high to fully cover the gap between the edges.
This high coverage regime forms GBs with spatially uncorrelated roughness, 
which quickly relax to smoother configurations.
Extrapolating the numerical results (with support from a theoretical approach)
to edge lengths and gap widths of some micrometers, we confirm
the advantage of this regime to produce GBs with minimal roughness faster
than near-equilibrium conditions.
This suggests an unexpected route towards efficient growth of two-dimensional materials with smooth GBs.
\end{abstract}


\section{Introduction}
\label{introduction}

The variety of applications of two-dimensional (2D) materials such as graphene and transition
metal dichalcogenides has motivated several studies of their growth kinetics in the last
two decades \cite{dongAM2019,huangNature2011,xuNatNano2016,taslim2019,ying2dMat2020}.
The growth of large grains is frequently desired to reduce the effects
of grain boundaries (GBs) on the electronic and mechanical properties.
However, in some materials, a large GB density may be a minor problem;
this is the case for the integration of ultrathin $\text{Mo}\text{S}_2$ films
with Si substrates in some devices that require low temperature deposition \cite{lin2dMat2021}.
Moreover, large GB densities may be beneficial for some applications of 2D materials,
such as $\text{W}\text{S}_2$ in sensors of Hg \cite{liuNatComm2021} and
$\text{Mo}\text{S}_2$ in the production of memtransistors \cite{sangwanNature2018}.
GBs of $\text{W}\text{S}_2$ may also be useful as high conductivity channels \cite{zhouACSAMI2019}.

Beyond the GB density, the GB roughness may also affect the
properties of 2D materials in nontrivial ways.
Sinusoidal graphene GBs are advantageous over atomically flat GBs because they
increase the mechanical strength and improve electronic properties \cite{zhangAFM2015}.
The fractures of $\text{Re}\text{S}_2$ monolayers more easily occur at GBs parallel
to Re chains \cite{zhangSmall2021}, so that disordered GBs may improve their properties.
In the case of graphene, there is also evidence that the grain size at the micrometer
scale does not have relevant effects on the electric conductivity \cite{ryuACSNano2014}
and that the mechanical strength is not reduced in the presence of GBs \cite{leeSci2013}.

The GB morphology is strongly related to the growth conditions
and several models were already proposed to explain this relation
\cite{dongAM2019,xuNatNano2016}.
For sufficiently slow graphene growth, the so-called attachment limited (or near-equilibrium)
conditions are observed, which usually lead to formation of smooth GBs.
Models of step flow describe the edge propagation \cite{artyukhovPNAS2012}
and the GB structure can be determined by the misorientation angle \cite{betsACSNano2021}
in these conditions.
Similar models and other multi-scale approaches are used to describe the growth of 2D
metal dichalcogenides \cite{chenACSAMI2019,momeni2020,hickeyNL2021}.
However, fast growth is frequently important for applications.
In this case, the so-called diffusion limited regime is frequently observed, in which
GBs significantly deviate from a straight profile.
Experiments, phase field models, and kinetic Monte Carlo (kMC) simulations show formation of grains
with fractal or dendritic morphologies that reproduce experimental observations
\cite{niePRB2011,mecaNL2013,wuPRL2015,xuNatNano2016,zhuangPCCP2016,yue2dMat2017,momeni2020}.

The recurrent observation of similar relations between the GB morphology and the growth conditions in 
a wide variety of materials is a motivation for the study of models that only contain the essential
physico-chemical mechanisms that are needed to explain those relations.
This reasoning has already been successful in the description of the morphology of 
homoepitaxial metal and semiconductor films by means of generic models including adatom diffusion
and attachment/detachment from islands and terraces
\cite{michely,pimpinelli,saitobook,etb,Misbah2010,einax}.

In this work, we introduce a model for the propagation of two grain edges 
and for the relaxation of the GB formed after the collision of those edges.
The aim of the model is
to identify the generic consequences of elementary microscopic
processes such as adatom surface diffusion, attachment, and detachment from edges and GBs
of a variety of 2d materials, in contrast with approaches that 
focus on particular applications
\cite{chenACSAMI2019,gaillard2015,enstone2016,betsACSNano2021}.
This modelling approach allows one to investigate the effects of the variations of
the rates of those processes over several orders of magnitude using
kinetic Monte Carlo (KMC) simulations.

The simulation results reproduce the main
features of the regimes limited by attachment and diffusion described above, respectively for low and high fluxes.
However, if the precursor flux is sufficiently high to produce  full coverage between the grain edges when they begin to grow, the onset of a nontrivial regime is observed:
after an uncorrelated growth of the edges,
relatively smoother GBs are formed at relatively shorter times compared to the other regimes.
From the simulation results at nanoscale and the support of a theoretical approach,
extrapolations to micrometer sized edges separated by micrometer sized gaps are performed.
They show the possible advantages of deposition with high coverages
for reducing GB roughness.

\section{Model and Methods}
\label{basic}

\subsection{Model for deposition, diffusion, and aggregation}
\label{modelsection}

The solid structure is modelled by a square lattice, which differs from the geometry of most 2D materials,
but is amenable to computational and analytical calculations.
The substrate below this square lattice is assumed to be inert.
Each lattice site may be empty or occupied by a particle, which may represent a graphene atom or a metal dichalcogenide molecule.
The lattice constant is denoted as $a$.

\begin{figure}[!ht]
\center
\includegraphics[clip,width=0.45\textwidth,angle=0]{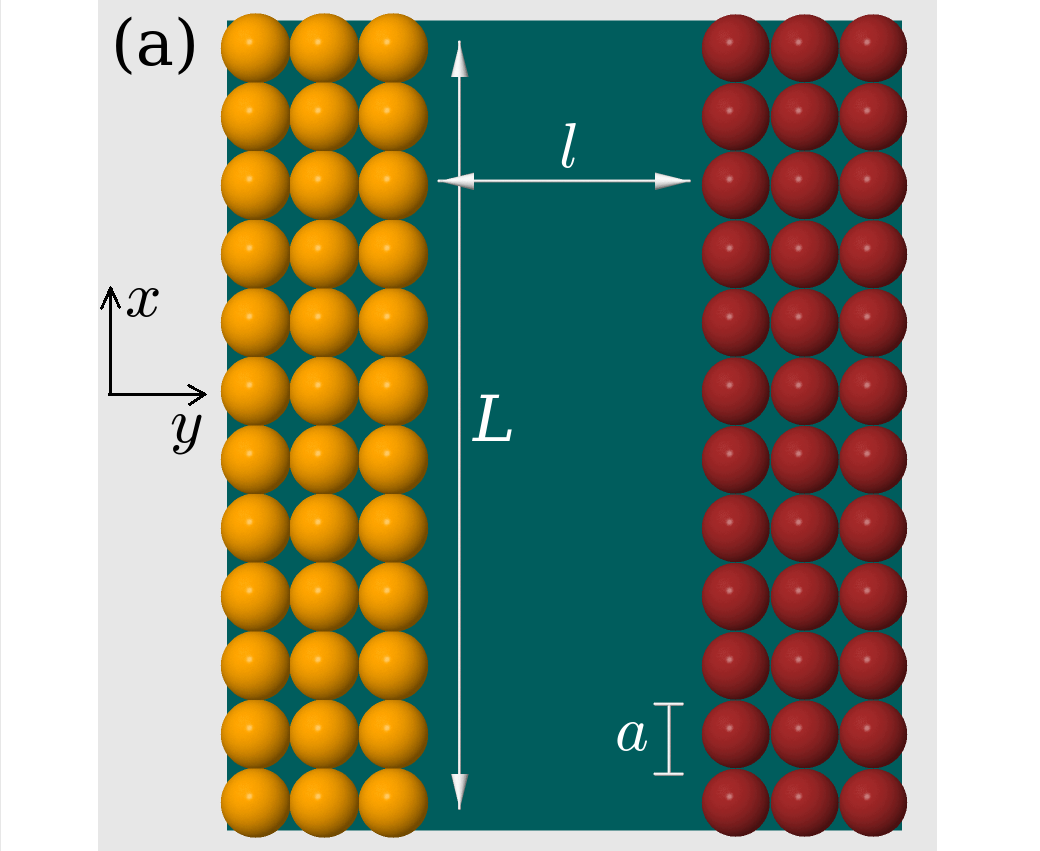}\\
\includegraphics[clip,width=0.45\textwidth,angle=0]{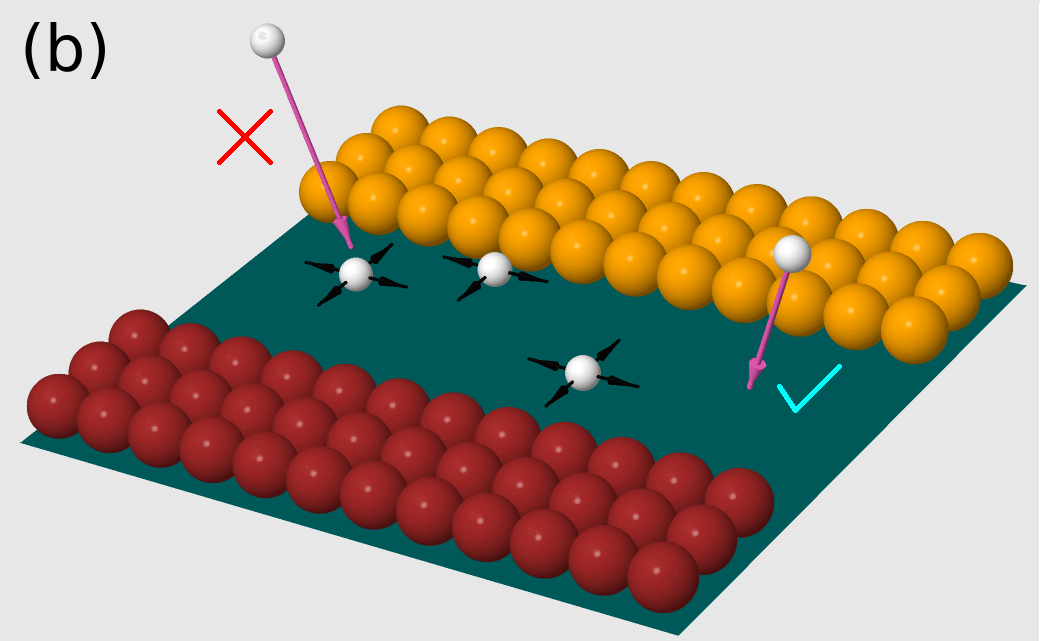}\\
\includegraphics[clip,width=0.45\textwidth,angle=0]{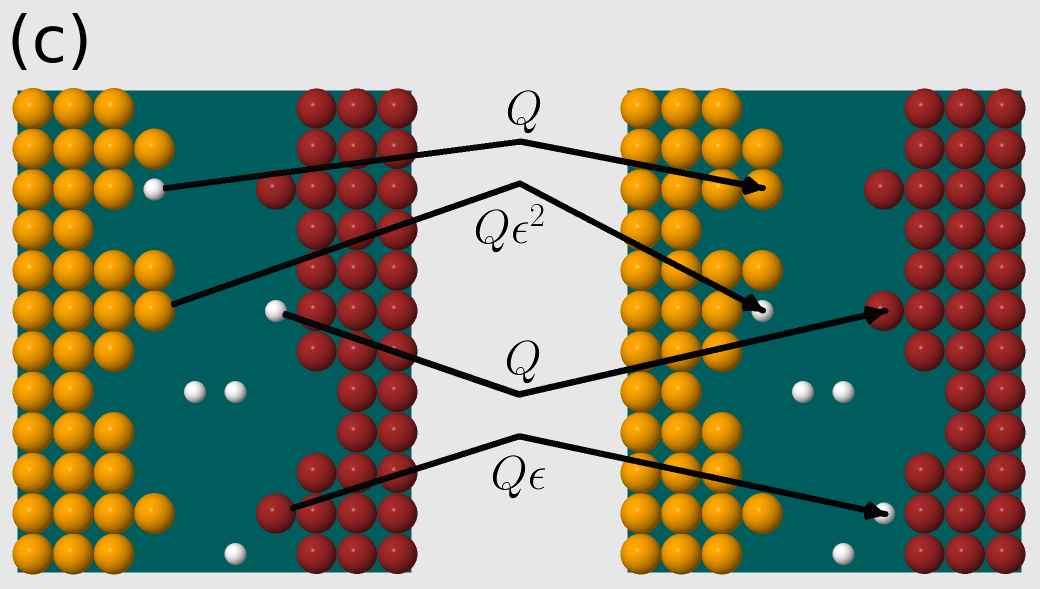}
\caption{
(a) Initial configuration of the grains A (yellow) and B (red).
(b) Two particles incide at the gap (magenta arrows), the right one is accepted and the left one is rejected (excluded volume).
Possible hops of particles already deposited in the gap are also indicated (black arrows).
(c) Rates of the processes that bring the configuration at the left to the configuration at the right.
}
\label{model}
\end{figure}

Two solid grains have initially flat interfaces (edges) of length $L$ along the $x$ direction
and are separated by a gap of width $l$,
as illustrated in Fig. \ref{model}(a).
Periodic boundaries are considered in the $x$ direction.
The model does not describe the process of grain nucleation and growth that lead to
this initial condition because its aim is to focus on the late stages of edge propagation
and on the coarsening of the GBs formed after their collision.
Once they are incorporated in one of the grains, the particles
are labelled with a index, A or B, that depends on the grain.
These indices are kept even after the complete filling of the gap,
so that the two grains do not coalesce into a single grain.
This approach does not reproduce the details of the anisotropy
and atomic rearrangements within the grain. However, it aims at catching
basic features of the growth of the edges, as well as the 
slow reshaping of the GBs after their collision.
These basic features should also be present in the case of grains with different crystallographic
orientations or with some mismatch that prevents coalescence; note that imperfect match
of 2D grains may also occur when they have the same orientation,
as observed in $\text{W}\text{S}_2$ growth on sapphire \cite{hickeyNL2021}.

The external particle flux $F$ is the number of incident particles per site per unit time.
If the incident particle reaches an empty site above the substrate, it becomes a mobile particle at that position; however, if it reaches an occupied site, the deposition attempt is rejected.
This process is illustrated in Fig. \ref{model}(b).

The mobile particles can diffuse on the substrate, with a hopping rate to
nearest neighbor (NN) sites per unit time equal to $D$.
A hop attempt is executed only if the target site is empty, otherwise the mobile particle
remains in the same position.
The process is also illustrated in Fig. \ref{model}(b).
The corresponding tracer diffusion coefficient is $Da^2/4$.
The assumption that the excluded volume is the only type of interaction between mobile
particles implies that other islands or grains are not formed in the gap.

The grains evolve via the attachment and detachment of particles.
To facilitate the connection with kinetic roughening theory and other analytical approaches,
these processes in each grain are restricted to the topmost point in each lateral coordinate $x$
(this is called the Solid-On-Solid constraint in the kMC and kinetic roughening literature).
When a mobile particle is at the site immediately above the topmost grain particle,
its attachment to the grain occurs with rate $Q$.
Simultaneously, a topmost particle can detach from a grain with rate $Q\epsilon^n$,
where $n$ is the number of NNs of that grain and $\epsilon<1$ is termed
the detachment probability per NN.
These processes are illustrated in Fig. \ref{model}(c).
With these rules, the attachment rate does not depend on the local configuration
of the grain edge.
However, the detachment is facilitated at edge tips ($n=1$) and is more difficult at
flat regions of the edges ($n=3$).

After the edge collision, the boundary between the grains A and B is formed.
The above rules for attachment
and detachment are maintained and
the interactions between particles of different grains are restricted to excluded volume interactions.
This means that the formation of bonds between NN particles of different grains is neglected.
This may be a reasonable approximation if the actual inter-grain bonds of a material
are sufficiently weak in comparison with the intra-grain bonds.
In this approximation, the characteristic time scales of the grain edge dynamics are not much
different before and after the collision.

The diffusion and detachment of particles are expected to be temperature
activated processes, so that the rates $D$ and $\epsilon$ increase as the temperature increases.
In many systems, the attachment is also activated, so $Q$ also increases with the temperature.
These parameters may then be written in Arrhenius forms 
(see Sec. S.I of the Supplementary Material).
It defines a
true thermodynamic equilibrium in the model without deposition,
which is enforced by detailed balance.
However, since our aim is to address possible properties of several 2D materials, we do not choose a set
of activation energies for the simulations, but study a large variety of possible relations
between those rates and the flux $F$.

\subsection{Simulations and quantities of interest}
\label{simulation}

Simulations were performed in lattices with edge length $L$ varying from $128a$ to $1024a$;
most results are obtained for $L=512a$.
The initial gap width $l$ varies between $16a$ and $128a$.
The model rates in the simulations are measured in terms of the flux rate $F$.
The ratios $F/D$ and $F/Q$ vary from ${10}^{-8}$ to ${10}^{4}$; these limiting values
respectively represent very low and very high fluxes for a given temperature.
Most simulations are performed with $\epsilon=0.1$, which implies that the detachment
rates have a weak dependence on the coordination number $n$; this is reasonable only
at high temperatures for most materials ($k_BT\approx 0.43$ times the activation energy
per NN).
Additional simulations were performed with $0.01\leq\epsilon\leq0.07$ to analyze the
effects of this parameter.

For most parameter sets, average quantities are calculated over $100$ different configurations,
but in some cases ${10}^3{\text{--}}2\times{10}^3$ configurations are considered.
The kinetic Monte Carlo algorithm developed for these simulations is
similar to that of previous works on submonolayer growth \cite{submonorev,Gagliardi2022,Saito2012}.

The main quantity calculated here is the roughness of the grain edges.
Letting $h\left( x,t\right)$ to denote the height of an edge at position $x$ and time $t$
measured relatively to its initial position, the roughness is defined as
\begin{equation}
W\equiv {\langle \;{\overline{{\left( h-\overline{h}\right)}^2}}\;\rangle}^{1/2} ,
\label{defW}
\end{equation}
where the overbars denote a spatial average (over the values of $x$ in a given sample)
and the angular brackets denote a configurational average (over different samples).
Due to the symmetry of the grains A and B, our configurational average also includes
averaging over the two grains of each growing sample.

We also measure the gap coverage, which is the fraction of the initial gap covered by deposited particles:
\begin{equation}
\theta = \frac{N_d}{\left( L/a\right)\left( l/a\right)} ,
\label{deftheta}
\end{equation}
where $N_d$ is the total number of deposited
particles.
Observe that particle attachment and detachment do
not affect this quantity.

To characterize the transition from edge growth (when attachment dominates over detachment) to GB relaxation
(when the grains evolve with balanced attachment and detachment),
we define an average collision time $t_c$.
For each position $x$ and a given configuration of evolving grains,
the local collision time $t_c^{\left( local \right)}\left( x\right)$ is defined as the first time in
which grains A and B occupy NN sites at position $x$
(with any value of $y$).
 Observe that the GB is not frozen after collision, so
particles may detach from
one of the grains, move to NN sites, or reattach to one of the
two grains.
The value of $ t_c$ is obtained by averaging $t_c^{\left( local \right)}$ over all positions $x$
and over different configurations of evolving grains; this averaging also
gives the standard deviation of the collision times, $\Delta t_c$.

\section{Results and Discussion}
\label{results}

\subsection{Morphological evolution}
\label{qualitative}

\begin{figure}
\center
\includegraphics[clip,width=\textwidth,angle=0]{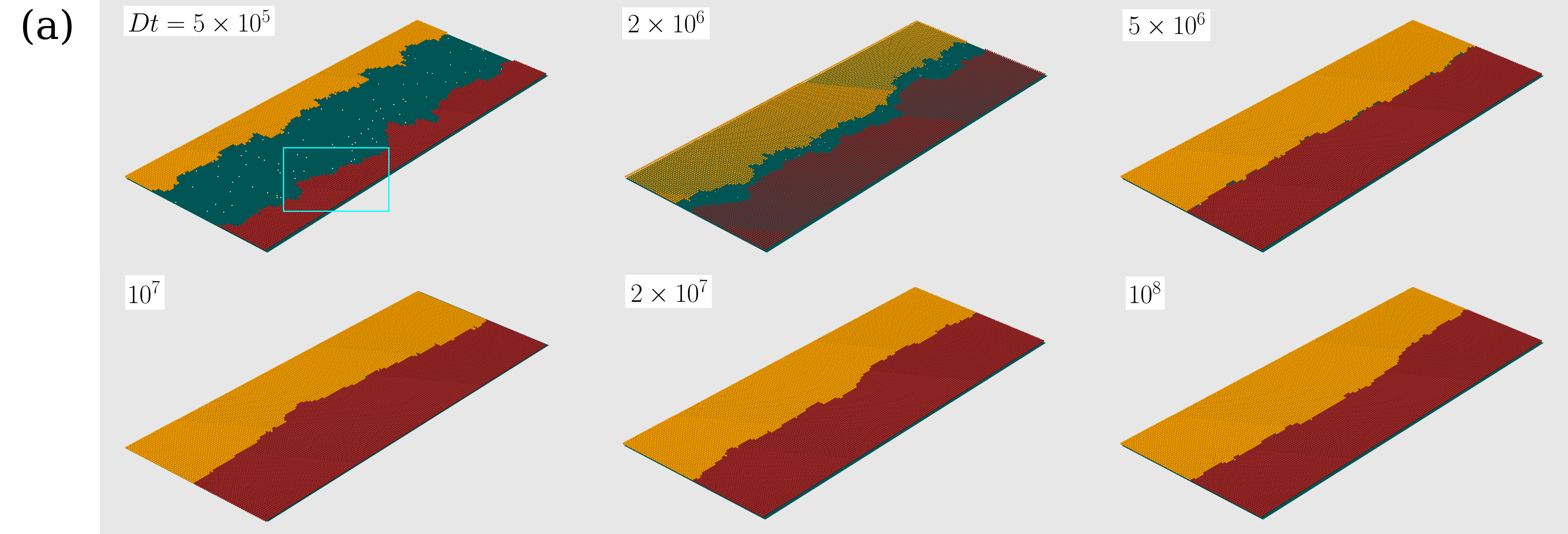}\\
\includegraphics[clip,width=\textwidth,angle=0]{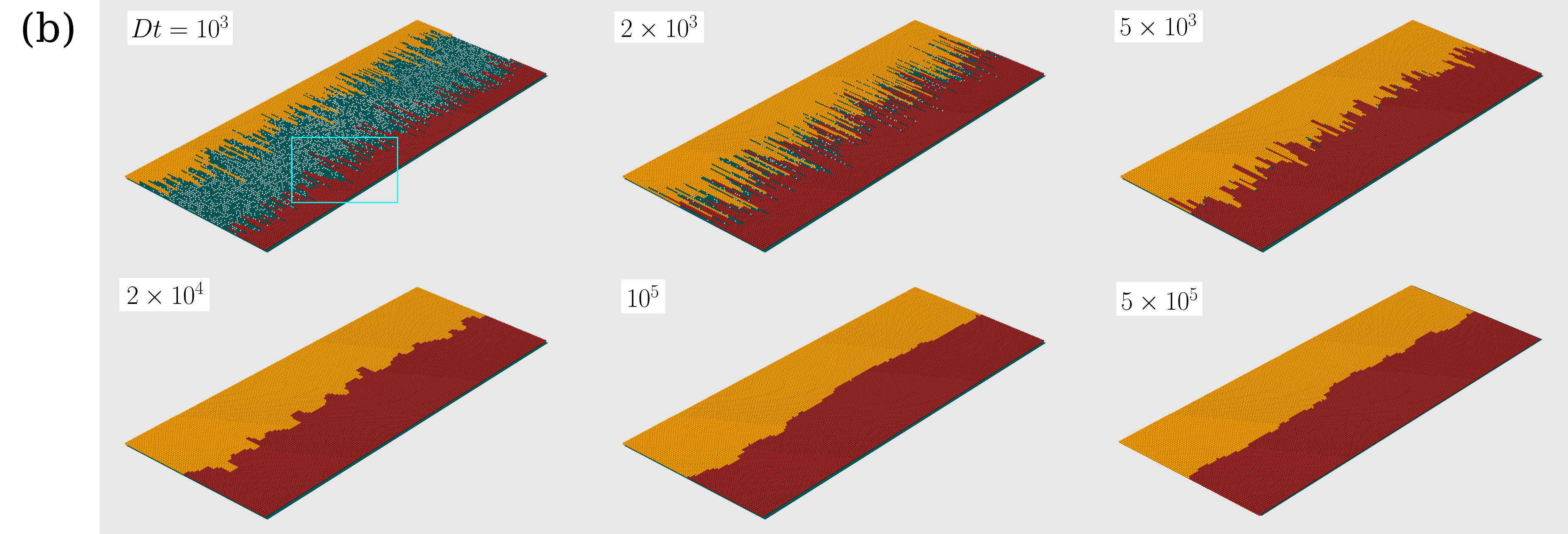}\\
\includegraphics[clip,width=\textwidth,angle=0]{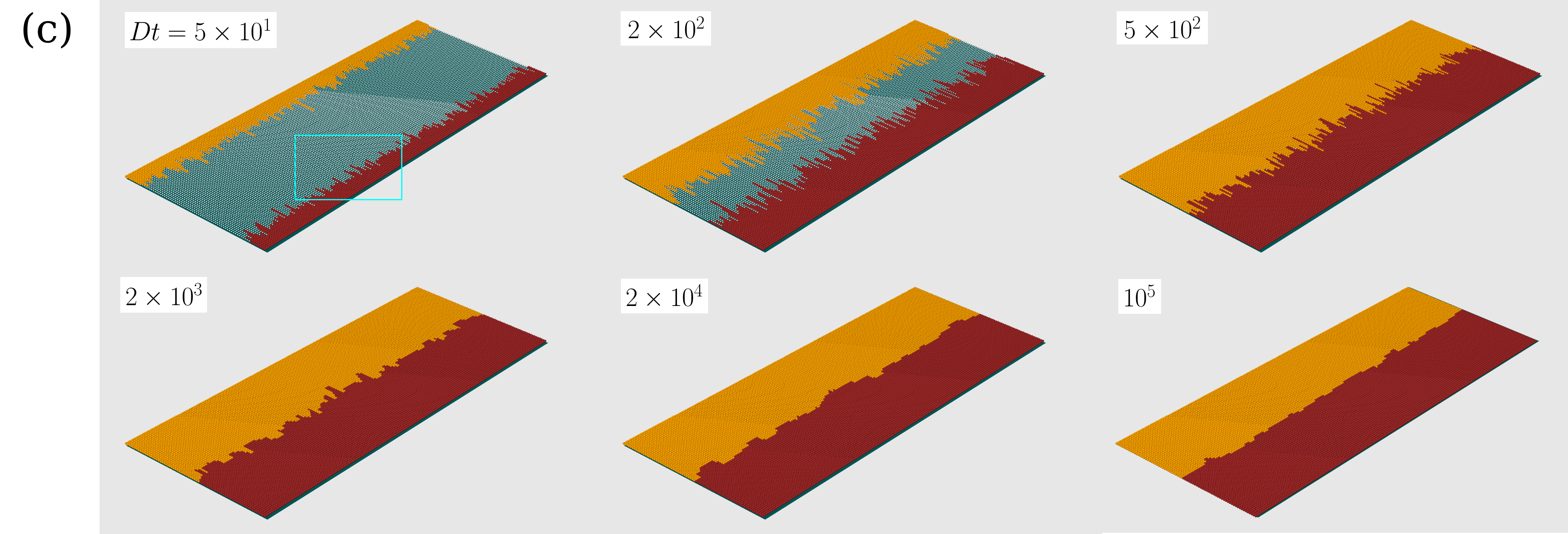}
\caption{
Snapshots of parts of the deposits grown with $Q/D={10}^{-1}$, $\epsilon=0.1$, $L=1024a$, and $l=128a$:
(a) $F/D={10}^{-6}$; (b) $F/D={10}^{-3}$; (c) $F/D=10$.
Light blue rectangles indicate the regions highlighted in Fig. \ref{zoom}.
}
\label{images}
\end{figure}

Figs. \ref{images}(a)-(c) show snapshots of the deposits grown with different values
of the incident flux but with constant values
of $D$, $Q$, and $\epsilon$, which are
set to $Q/D={10}^{-1}$ and $\epsilon=0.1$.
This is expected to mimic constant temperature conditions with variable precursor fluxes.
The growth times are shown in dimensionless units of $Dt$.

\begin{figure}[!ht]
\center
\includegraphics[clip,width=0.8\textwidth,angle=0]{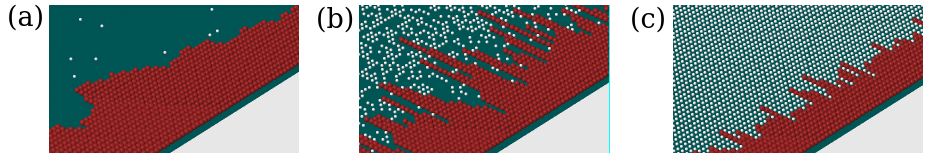}
\caption{
(a), (b), and (c) show zooms of the regions indicated in Figs. \ref{images}(a), \ref{images}(b),
and \ref{images}(c), respectively.
}
\label{zoom}
\end{figure}

For the lowest flux ($F/D={10}^{-6}$), Fig. \ref{images}(a) shows that
the roughness of the growing edges is dominated by long wavelength fluctuations.
The mounds and valleys evolve from typical sizes $10a{\text{--}}30a$ at $Dt=5\times{10}^5$
to sizes larger than $30a$ at $Dt=2\times{10}^6$, but no significant change in the
edge roughness is visible.
The particle density in the gap is very small until the grain collision;
this is confirmed by the magnified view in Fig. \ref{zoom}(a) of the region highlighted
in Fig. \ref{images}(a).
Similar evolution is observed for larger diffusion coefficients of the particles in the gap,
i.e. when $Q/D$ and $F/D$ decrease by the same factor, as shown in Sec. S.II
of the Supplementary Material.
These features are representative of the attachment limited growth regime in this model.

As the gap is closed and a GB is formed, there is a drastic decrease of the
boundary roughness.
This decrease is also observed in models of collisions of uncorrelated interfaces
that propagate with finite velocities and is independent of the detailed physics
of their short range interactions \cite{reispierrelouis2018}.
After the collision, the GB evolves slowly by detachment of particles from
one edge followed by reattachment to the other, with apparently small changes in the roughness.

For the intermediate value of the flux ($F/D={10}^{-3}$), but the same ratio $Q/D$,
Fig. \ref{images}(b) shows spikes at the grain edges since early times.
At $Dt={10}^3$, a high particle density is already present in the gap. note that this time is
much shorter than that of the first panel of Fig. \ref{images}(a).
The magnified view in Fig. \ref{zoom}(b) shows that the particle density is depleted only
near the valleys of the grain edges.
For this reason, the spikes grow faster than the valleys, as noticeable by comparing 
the snapshots at $Dt={10}^3$ and $2\times{10}^3$ in Fig. \ref{images}(b).
This is characteristic of unstable growth in a diffusion-limited regime.

The spikes that result from the unstable growth
are particular features of this model.
They contrast with the fractal
or dendritic shapes observed in experiments and in other models
with unstable growth \cite{xuNatNano2016,momeni2020}.
This difference is a consequence of the condition that particle attachment
occurs only at the topmost site at each $x$, which prevents the lateral attachment
responsible for more complex morphologies.

When the gap is completely filled, the spikes remain, so the GB roughness is initially large.
At $Dt=2\times{10}^4$, Fig. \ref{images}(b) shows that the spikes are transformed to shorter
and thicker structures.
A smooth GB is achieved at $Dt={10}^5$, which is a much shorter time than that of
the attachment-limited mode shown in Fig. \ref{images}(a).

Fig. \ref{images}(c) shows results for much higher flux, $F/D=10$, again with the same $Q/D$.
Now the particle attachment is very slow compared to the incoming flux.
At a short time, $Dt=5\times{10}^1$, the gap is fully covered
[see the magnified view in Fig. \ref{zoom}(c)] and so it remains until the two edges
collide and form a GB.
This contrasts with the diffusion-limited regime, in which the density was depleted
near the edge valleys.
In the present situation, the mobile particles are effectively static.
The attachment to the grain edges is consequently uncorrelated since short times because
all points of the edge (at valleys or crests) are in contact with one of those particles.
Fig. \ref{images}(c) also shows edges with spikes, but they are thinner and less
protuberant than those of the diffusion-limited regime [Fig. \ref{images}(b)].

The GB is formed at $Dt\sim{10}^3$ with a non-negligible roughness.
The relaxation of the GB leads to its smoothening.
This process is comparably faster than that reported in Fig. \ref{images}(b), although the only relevant parameters, $Q$ and $\epsilon$, are the same.
Indeed, at $Dt=2\times{10}^4$, the surface roughness 
in Fig. \ref{images}(c) is much smaller than that observed
in Fig. \ref{images}(b).

In Sec. S.II of the Supplementary Material, we show that similar morphologies are
obtained if the ratio $Q/D$ decreases by a factor ${10}^2$, corresponding to a
larger particle diffusivity.
A full coverage of the gap is also attained at short times
for a high value of the flux, which corresponds to the
same ratio $F/Q={10}^2$ of Fig. \ref{images}(c).

Thus, an apparently different growth regime is observed when the precursor flux is sufficiently high to produce full surface coverage
before a significant edge displacement.
In this high coverage regime, the GB may have an initially large roughness,
but it attains smooth configurations in shorter times than those observed in diffusion-limited and attachment-limited regimes.

\subsection{Roughness, coverage, and edge collision time}
\label{roughnessevolution}

\begin{figure}[!ht]
\center
\includegraphics[clip,width=\textwidth,angle=0]{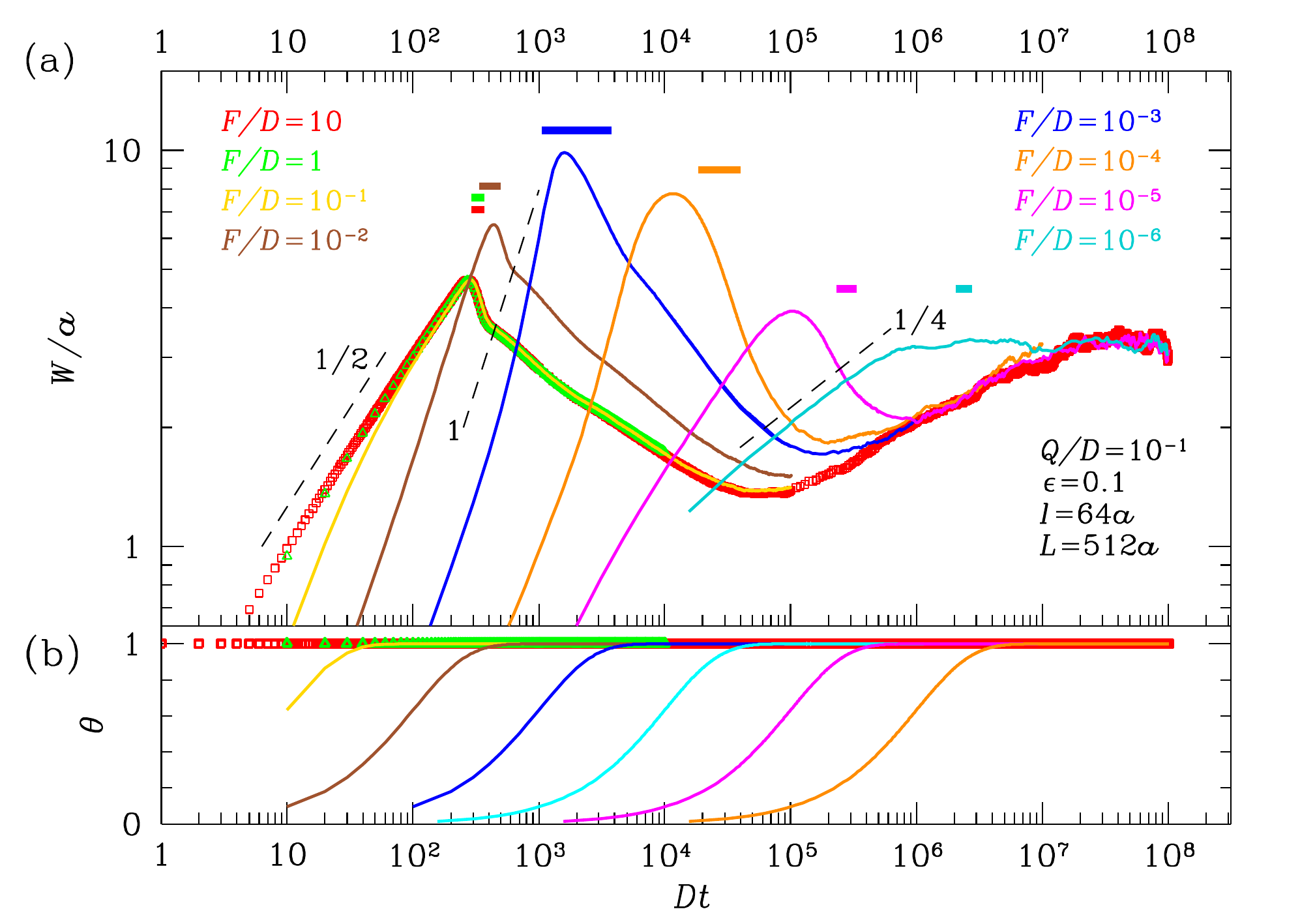}\\
\caption{
(a) Roughness evolution for constant $Q/D={10}^{-1}$ and variable flux.
Dashed lines with the indicated slopes are drawn for comparison with the slopes of the plots.
Horizontal bars at $W/a\lesssim1$ indicate the edge collision times
(average plus and minus one standard deviation).
(b) Evolution of the gap coverage for the same
parameter sets, with the same color code of (a).
}
\label{roughness}
\end{figure}

Fig. \ref{roughness}(a) shows the time evolution of the roughness $W$ of 
the grain edges for $Q/D={10}^{-1}$, $\epsilon=0.1$, initial gap width $l=64a$,
edge length $L=512a$, and several values
of $F/D$.
For most of the flux values, Fig. \ref{roughness}(a) also indicates
the typical ranges of collision times, which indicate the GB formation.
Fig. \ref{roughness}(b) shows the coverage $\theta$ as a function of time for the same
parameter sets.

In all cases, $W$ increases 
during edge growth,
as expected for any kinetic roughening process starting from a flat configuration
\cite{barabasi,krug}.
In most cases, $W$ attains a maximal value near the average collision time
(the only exception is the case with the lowest flux, in which $W$ seems to
saturate after the initial increase).
That result parallels the previous observation of a decrease of the roughness during
the collision of interfaces propagating with constant velocities, independently
of their short range interactions \cite{reispierrelouis2018}.
The smoothening or roughening processes that take place after the formation
of the GB depend on the flux.

The lowest flux in Fig. \ref{roughness}(a), $F/D={10}^{-6}$, 
corresponds to the snapshots in Fig. \ref{images}(a). The plot shows that
the initial near-equilibrium growth occurs with $W\sim t^{1/4}$.
This is the roughening described by the Edwards-Wilkinson (EW) equation
\cite{ew} at coarse grained scale, in which the surface tension and an uncorrelated noise
are the main mechanisms.
The relative fluctuation of the collision time is not large, as inferred by the
width of the indicative bar in Fig. \ref{roughness}(a).
After the GB formation, there are small changes in the roughness, which indicate that
the GB attains a stationary regime.

The stationary regime can actually be mapped to a true thermodynamic equilibrium state,
in which the roughness depends only on the detachment probability
$\epsilon$ and on the edge length $L$;
see details in Sec. S.III of the Supplementary Material.
For all values of the model rates ($F$, $D$, and $Q$) and of the initial gap width,
the roughness is expected to converge to the same stationary value at sufficiently long times.
Thus, depending on the maximal roughness attained before the grain collision,
$W$ may increase or decrease after formation of the GB.
For instance, Fig. \ref{roughness}(a) shows that  $W$ increases towards the stationary value
for $F/D={10}^{-5}$ and $F/D={10}^{-4}$;
however, Sec. S.IV of the Supplementary Material shows that $W$
decreases towards the stationary value when the gap width is smaller.

For the intermediate flux, $F/D={10}^{-3}$, Fig. \ref{roughness}(a) shows that
the initial increase of the roughness is faster than linear;
it corresponds to the snapshots in Fig. \ref{images}(b).
The explosive increase of height fluctuations confirms the unstable growth which is
expected in a diffusion-limited regime.
The range of collision times is broader in this case, as a consequence of the 
large dispersion in the local heights of the two edges.
When the edges collide,
$W$ decreases and can reach much smaller values as the GB relaxes.
After reaching a minimal value, $W$ increases again
(it is expected to converge to the stationary value at long times).
In previous studies of sudden changes from a first kinetics that
produces faster roughening to a second kinetics that produces slower
roughening, similar results were obtained,
viz. rapid smoothening after the change and subsequent slow roughening
\cite{assisreis2014,smoothening2015}.

For the largest flux ($F/D=10$),
the roughness in Fig. \ref{roughness}(a) approximately
increases as $W\sim t^{1/2}$;
the corresponding snapshots are in Fig. \ref{images}(c).
This is consistent with random uncorrelated
deposition at both grain edges \cite{barabasi}.
Fig. \ref{roughness}(b) shows that full substrate coverage is attained
long before the collision time, in contrast with
the other regimes, in which the coverage is $1$ only at times near the collision.
This confirms that the separated grain edges grow
with a fully covered gap.
The maximal $W$ is also attained as the edges collide and the subsequent
decrease is associated with the GB formation.
However, the range of collision times is small compared to the unstable regime.
After the GB is formed, the roughness attains a minimum value which is smaller
than the values obtained
in the near-equilibrium and in the diffusion-limited regimes.
Moreover, this mininum value is attained
in a shorter time compared to those regimes,
as previously suggested by the snapshots of Figs. \ref{images}(a)-(c).

A common feature of the regimes limited by attachment and by diffusion
is that the time of GB formation (the average collision time) is roughly
proportional to the reciprocal of the flux rate:
$t_c\sim3/F$.
The initial gap width and the mechanisms of particle diffusion, attachment, and detachment
determine the edge morphology.
However, in the high coverage regime (obtained with the highest fluxes),
the time of GB formation depends on the rate $Q$ but not on the flux $F$; 
for instance, compare the data for
$F/D=1$ and $F/D=10$ in Fig. \ref{roughness}(a).
After the gap is (rapidly) filled, the attachment rate $Q$ controls the edge propagation,
with negligible effect of the diffusion rate because the particle positions are effectively
frozen.
Thus, the collision time is inversely proportional to $Q$ and proportional to the gap width $l$;
see details in Sec. S.V of the Supplementary Material.

Between the three regimes described above, the roughness evolution shows crossover
features.
In Fig. \ref{roughness}(a), $F/D={10}^{-5}$ and ${10}^{-4}$
represent the crossover between the attachment-limited and the diffusion-limited
regimes, whereas $F/D={10}^{-2}$ and ${10}^{-1}$ represent the crossover
from the latter to the high coverage regime.

\begin{figure}[!ht]
\center
\includegraphics[clip,width=0.95\textwidth,angle=0]{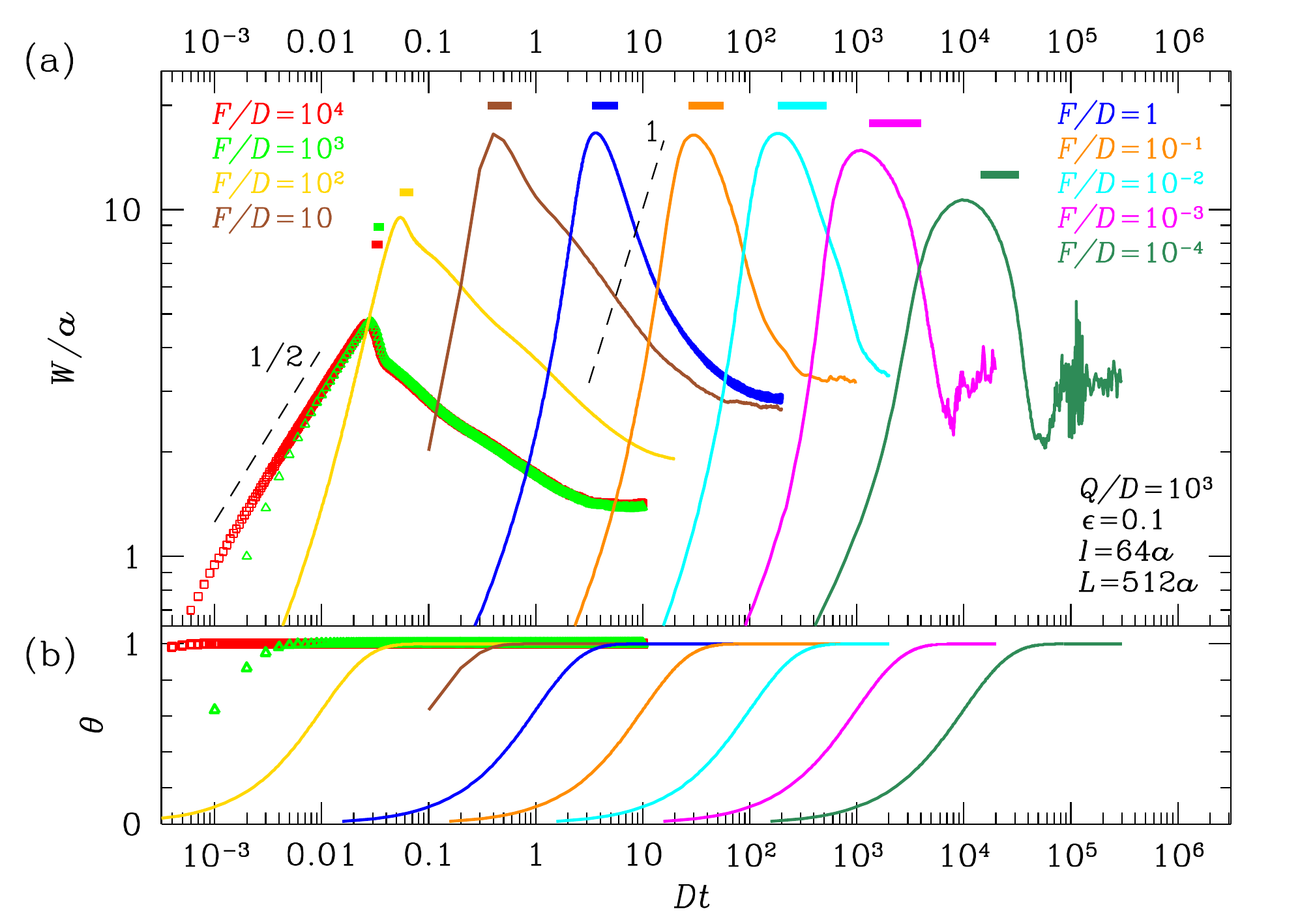}\\
\includegraphics[clip,width=0.95\textwidth,angle=0]{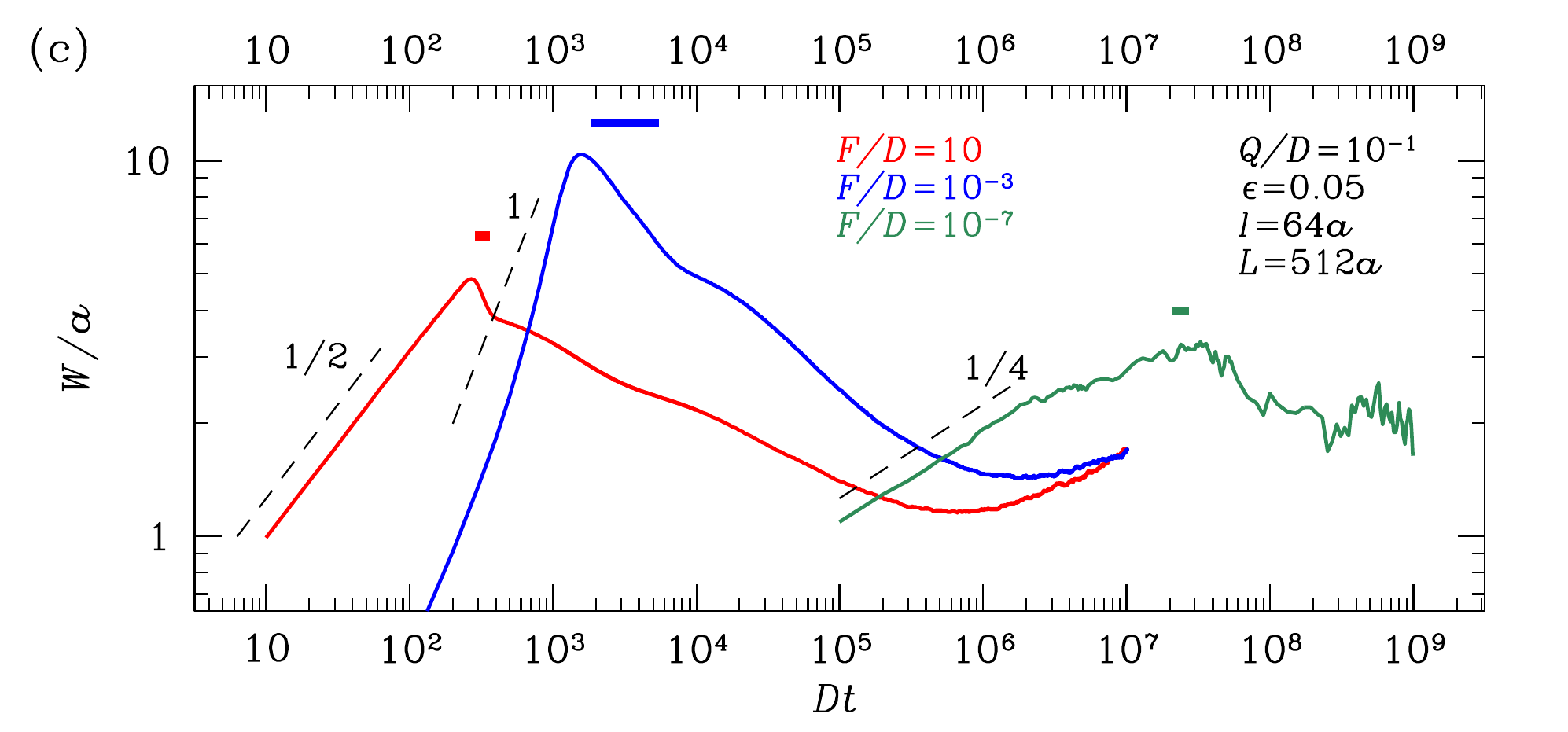}
\caption{
(a), (c) Roughness evolution for constant $Q/D$ and variable flux.
Dashed lines with the indicated slopes are drawn for comparison with the slopes of the plots.
Horizontal bars at $W/a\lesssim1$ indicate the edge collision times
(average plus and minus one standard deviation).
(b) Coverage evolution for the same parameters in (a), with the same color code.
}
\label{roughplus}
\end{figure}

Qualitatively similar results are obtained for other ratios $Q/D$. 
Fig. \ref{roughplus}(a) shows the roughness evolution for several values of the flux $F$
when $Q/D={10}^{3}$, which represents systems with very slow surface diffusion in
comparison with the attachment and detachment rates; the other parameters remain the
same as in Fig. \ref{roughness}(a) and the time scale of diffusion is also used
for the data presentation.
The main difference here is that the diffusion-limited (unstable) regime spans
a broader range of $F$, since the slow diffusion favors the particle
depletion near the edge valleys.
For this reason, the attachment-limited regime
is not observed in Fig. \ref{roughplus}(a).
However, or $F/D={10}^4$ (corresponding to $F/Q=10$),
the features of the high coverage regime are also observed:
the edges initially show the $W\sim t^{1/2}$ scaling of uncorrelated growth,
while the coverage has already reached the full
value $1$, as shown in Fig. \ref{roughplus}(b);
after the GB formation, $W$ reaches a minimal
value smaller than those attained in the other regimes, and this occurs at a shorter time.

Much larger diffusion coefficients (compared to the attachment rate) facilitate
the particle redistribution in the gap and permit the observation of the 
attachment-limited regime, but suppress the unstable regime (diffusion-limited).
However, for the largest flux rates, the features of
the high coverage regime are still observed because
the diffusion is not effective in this case.
This is illustrated in Sec. S.VI of the Supplementary Material for $Q/D={10}^{-3}$.

Changes in the gap width $l$ affect the crossover between the diffusion-limited 
and the attachment-limited regimes, but do not change the features of the very
high flux regime.
For smaller $l$, the unstable growth may not have sufficient time to develop, so
the EW roughening of the edges is observed for slightly larger fluxes; see Sec. S.IV of the Supplementary Material.
Instead, for larger $l$, the unstable regime extends to smaller fluxes because
the instability has longer time to develop before the edges collide.

Changes in the detachment probability $\epsilon$ (which relates the detachment rates
with the number of NNs) lead to some relevant changes in the roughness evolution, as shown in 
Fig. \ref{roughplus}(c) for $\epsilon=0.05$
and the other parameters kept the same as in
Fig. \ref{roughness}(a).
First, a smaller $\epsilon$ pushes the diffusion-limited regime to lower fluxes
because the detachment kinetics of the edges is slower, but the attachment does not change.
For this reason, a clear EW scaling of the roughness
is observed only for $F/D={10}^{-7}$
[in contrast with $F/D={10}^{-6}$ for $\epsilon=0.1$; Fig. \ref{roughness}(a) ].
Second, for the intermediate and the largest fluxes,
the smaller detachment rates leads to slower
decays of the roughness after the GB formation.
Despíte these differences, the high coverage regime obtained
with the highest flux ($F/D=10$) still provides the smallest
GB roughness during the relaxation.

For the largest flux in Fig. \ref{roughplus}(c) ($F/D=10$), a nontrivial
feature is the appearance of shoulders in the decay of the GB roughness.
The decrease of the roughness upon the collision is a stochastic effect characteristic of
interface collisions \cite{reispierrelouis2018}.
After that, the roughness decreases by the elimination of the 
thinnest spikes of the GB, which
requires the detachment of particles with
$n=1$ NN, whose rate is $Q\epsilon$.
For small $\epsilon$, this rate is much smaller than that of the initial growth ($Q$);
the first shoulder is a consequence of the longer time scale associated to this process.
Further relaxation of the GB depends on kink detachment
(see Sec. S.III of the Supplementary Material), whose rate is $Q\epsilon^2$.
This is smaller than the rate of spike elimination by the factor $\epsilon$,
which explains the second shoulder.
These features become more pronounced for smaller $\epsilon$, as shown in Sec. S.VII
of the Supplementary Material;
however, the high coverage
regime still provides the minimal roughness in a shorter time
than the other growth regimes.

\subsection{Phase diagram of grain edge growth}
\label{phasediagram}

\begin{figure}[!ht]
\center
\includegraphics[clip,width=0.45\textwidth,angle=0]{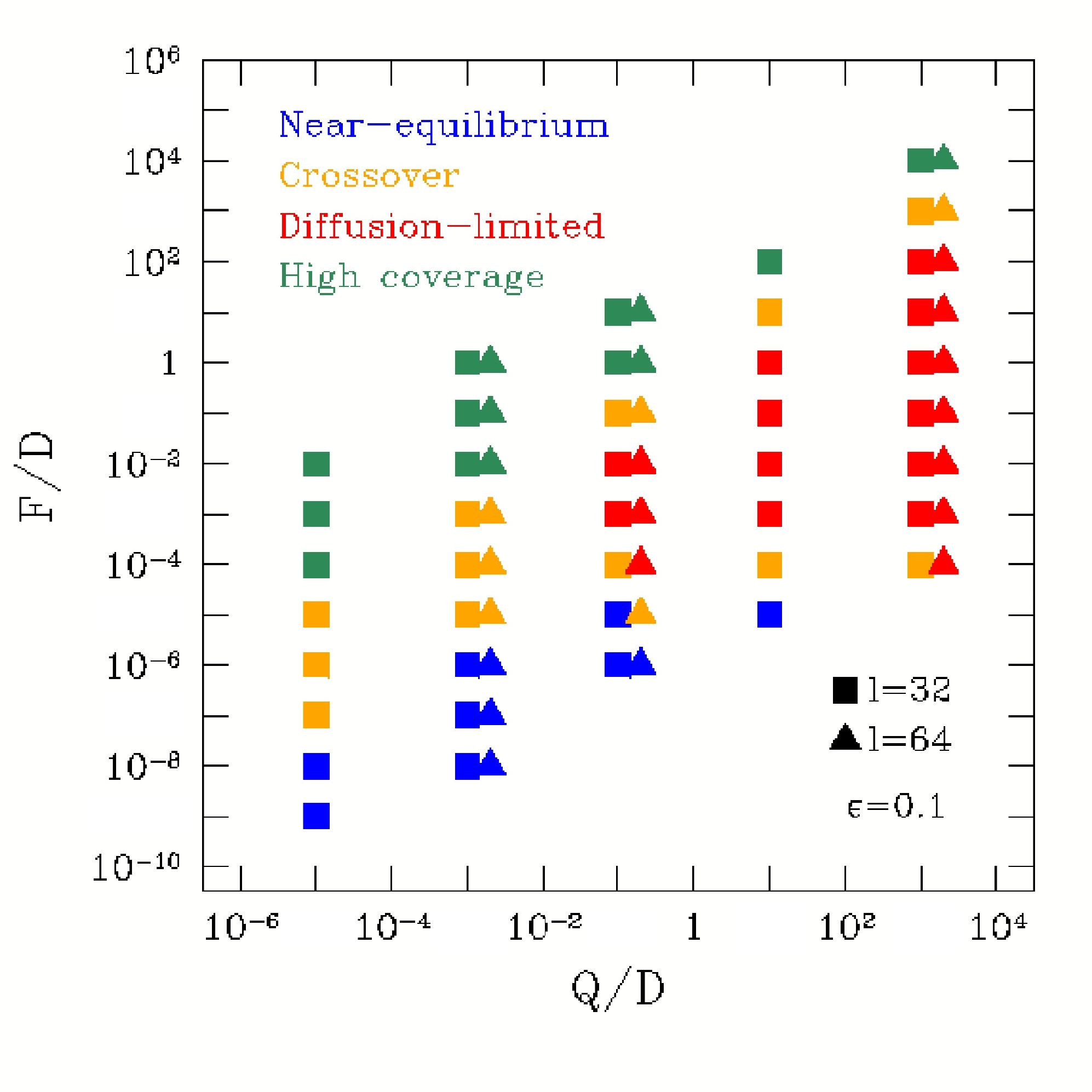}
\caption{
Grain growth regimes as function of the kinetic parameters for $\epsilon=0.1$ and two gap widths.
}
\label{diagram}
\end{figure}

In our simulations, we identify three regimes of edge growth, i.e. before the edges begin to
interact.
The attachment-limited or near-equilibrium regime is charaterized by slopes of the
$\log{W}\times\log{\left( Dt\right)}$ plots near the EW value $1/4$.
The diffusion-limited or unstable regime is characterized by slopes larger than $1$ at some time interval before the maximal $W$.
Finally, the very high flux or random growth regime is characterized by slopes near $1/2$.
Other behaviors are considered as crossovers.
Fig. \ref{diagram} shows the phase diagram obtained with these criteria for $\epsilon=0.1$
and gap widths $l=32a$ and $64a$.
For 2D materials such as graphene or metal dichalcogenides, the lattice constant is  $\sim0.3$~nm, so those values correspond to widths $\sim10{\text{--}}20$~nm.

For very slow attachment-detachment kinetics in comparison with diffusion (very small $Q/D$),
the unstable growth is not observed because the mobile particles are uniformly distributed
in the gap.
The unstable growth is typically observed for $F/Q\lesssim0.1$
(to avoid the rapid uncorrelated growth),
$F/D\gtrsim{10}^{-3}$ (to avoid very slow deposition, which favors near-equilibrium growth),
and $Q/D\gtrsim1$ (fast interface kinetics compared to diffusion).

However, these conditions change as the gap width $l$ or the detachment probability $\epsilon$ change.
For wider gaps, the instability has longer time to develop before the edge collision,
so it can be observed for smaller values of $F/D$.
As $\epsilon$ decreases, particle detachment becomes slower, which also favors the instabilities;
thus, the unstable regime can also be observed
for smaller values of $F/D$.

The high coverage regime is more robust against changes in the physical or
chemical parameters.
It typically appears for $F/Q\gtrsim10$, which is a condition in which the gap is almost
completely filled before the attachment of a single layer of particles at each edge.
As explained before, the diffusion coefficient $D$ is not important in this regime
because the particles in the gap cannot move to the occupied NNs.

\subsection{The minimal GB roughness}
\label{minimization}

In the simulations of growth with high precursor coverage ($F/Q\gtrsim10$),
we measured the minimal value of the roughness obtained during the GB
relaxation, $W_{min}$, and the time in which this minimum was attained, $t_{min}$.

\begin{figure}[!ht]
\center
\includegraphics[clip,width=0.45\textwidth,angle=0]{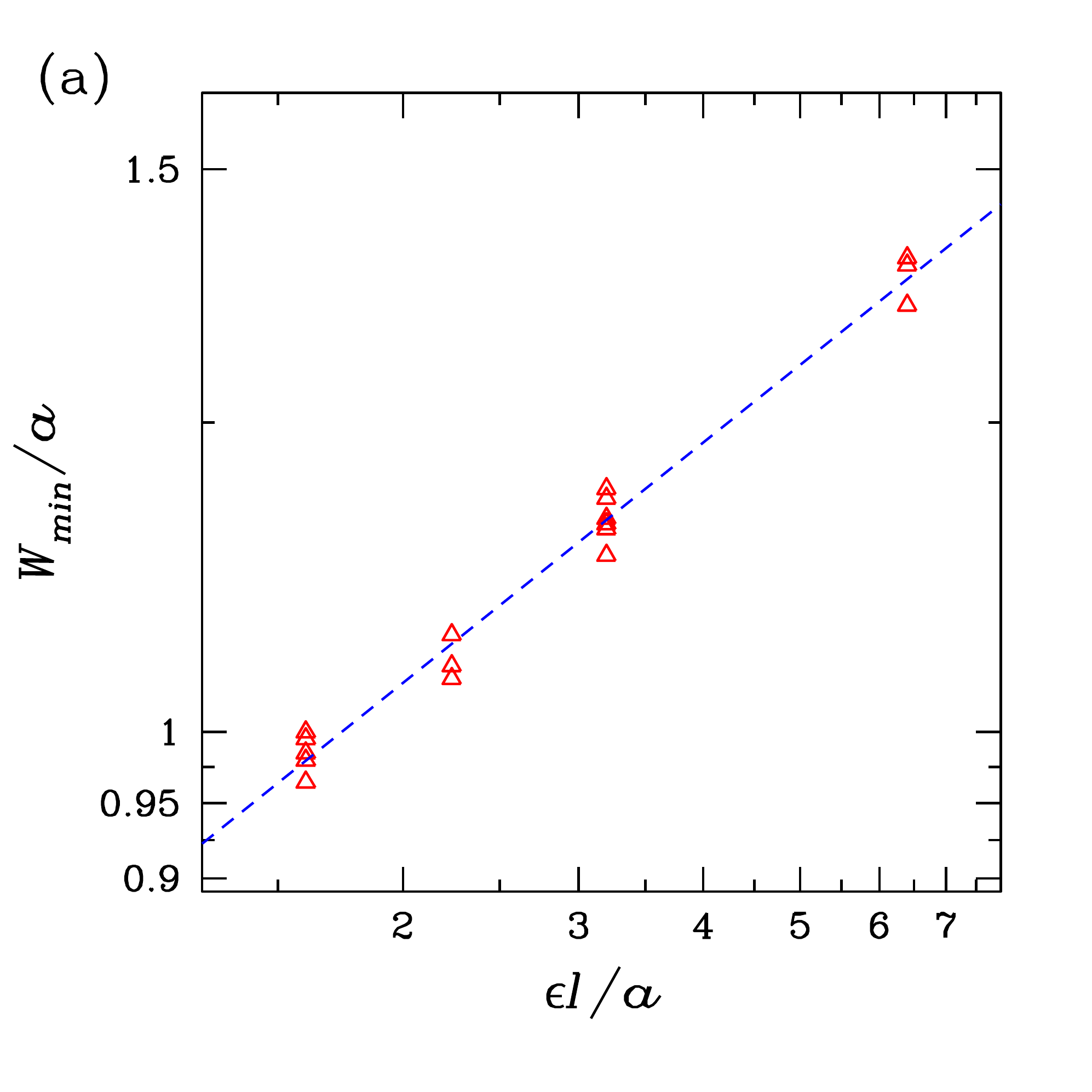}\\
\includegraphics[clip,width=0.45\textwidth,angle=0]{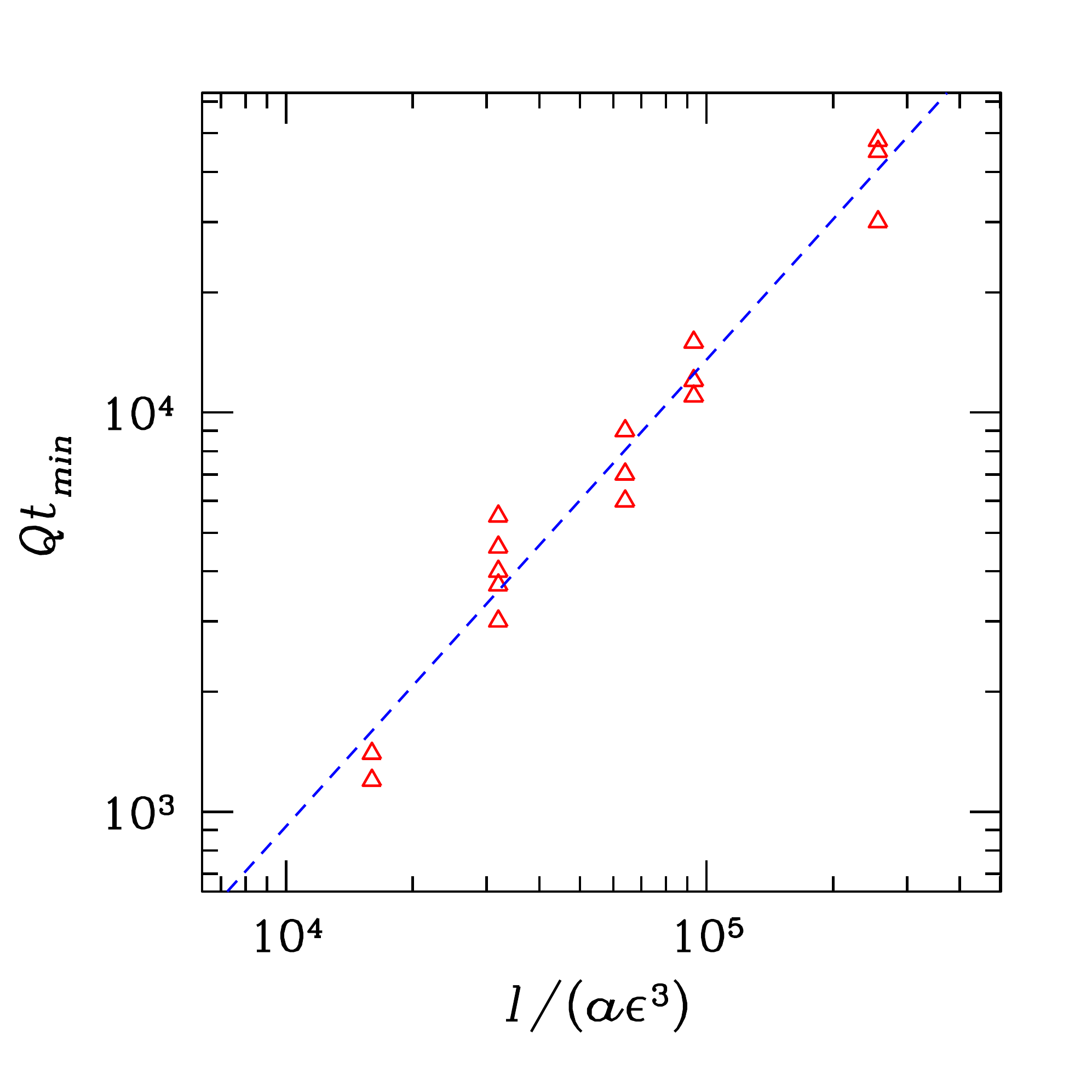}
\caption{
(a) Minimal roughness in the high coverage regime
as a function of $\epsilon l/a$.
The dashed line is a least squares fit with slope $0.25$.
(b) Scaled time of the minimal roughness as a function of $l/\left(a\epsilon^3\right)$.
The dashed line is a least squares fit with slope $0.16$.
}
\label{Wmin}
\end{figure}

$W_{min}$ does not depend on the parameters
$F$, $Q$, $D$, and $L$; it is affected only by the detachment rate
$\epsilon$ and the gap width $l$.
Fig. \ref{Wmin}(a) shows $W_{min}/a$ as a function of the scaling variable
$\epsilon l/a$, i.e. the dimensionless product of the above parameters.
Since the time of the minimal roughness rapidly increases as $\epsilon$ decreases and 
the simulations become slower as $l$ increases,
we considered restricted ranges for those variables
($0.05\leq\epsilon\leq0.1$ and $16\leq l\leq64$) to obtain accurate estimates of $W_{min}$.
The linear fit in Fig. \ref{Wmin}(a) gives the relation
\begin{equation}
W_{min} \approx 0.87{\left(\epsilon a^3 l\right)}^{0.25} .
\label{Wminscaling}
\end{equation}

The dependence of $W_{min}$ on $l$ can be explained by simple scaling arguments
\cite{assisreis2014,smoothening2015}.
From the initial flat edge condition to the edge collision,
each of the grains is displaced by an average length $l/2$;
thus, the average number of particles attached to each column $x$ of a grain is $N=(l/2)/a$.
The uncorrelated growth in this regime leads to a maximal roughness of order \cite{barabasi}
\begin{equation}
W_{max}\sim a\sqrt{N}=\sqrt{\frac{al}{2}} ;
\label{Wmax}    
\end{equation}
the characteristic rate of this growth is $Q$.
As the GB is formed, there is a sudden change from this uncorrelated growth to a correlated kinetics.
If the characteristic rate of attachment and detachment of the final kinetics is
the same as the initial one,
the uncorrelated fluctuations are rapidly suppressed
and the roughness  rapidly reaches the same time evolution of the final kinetics starting from a flat interface
(see e.g. Fig. 6 of Ref. \protect\cite{assisreis2014});
here, this would be the case of $\epsilon\sim1$.
The GB kinetics is EW (Sec. S.III of the Supplementary Material),
in which the growth from a flat interface leads to $W\sim aN^{1/4}$,
where a prefactor of order $1$ is excluded \cite{ew,barabasi}.
The minimal roughness is then obtained by substituting the value of $N$ at the collision
time in this relation, which gives $W_{min}\sim l^{1/4}$.

However, to describe the GB relaxation with smaller $\epsilon$, a more powerful approach has to be used.
We developed a Langevin model \cite{PierreLouis1998}
which assumes that the GB evolution is controlled by kink site detachment
from one grain (with characteristic rate $Q\epsilon^2$)
and attachment to the other grain.
In Sec. S.III of the Supplementary Material, this model is solved for an initial surface
configuration with uncorrelated roughness $W_{max}$ and predicts
\begin{equation}
W_{min}^{\left(\text{Lang}\right)}={\left[\frac{1}{\pi(1-\epsilon)^2}\right]}^{1/4}
{\left(a^3l\epsilon\right)}^{1/4} .
\label{WminL}
\end{equation}
Since $1-\epsilon\approx 1$ is a reasonable approximation in our simulations, Eq. (\ref{WminL})
leads to the same scaling in $l$ and $\epsilon$ of Eq. (\ref{Wminscaling});
the approximation gives a prefactor $0.75$,
which is not distant from the numerical estimate $0.87$
in Eq. (\ref{Wminscaling}).

The Langevin approach also predicts that the minimum roughness is attained at
\begin{equation}
    t_{min}^{\left(\text{Lang}\right)}=\frac{l}{2aQ}\left[1+
    \frac{(1-\epsilon^2)}{8\epsilon^3}\right] ,
\label{tminL}
\end{equation}
where the first term accounts for the uncorrelated growth of the edges of independent grains
and the second one accounts for the coarsening of the GB.
For $\epsilon\ll1$, the first term is negligible in comparison with the second one,
which gives $t_{min}^{\left(\text{Lang}\right)}\approx l/\left( 16aQ\epsilon^3\right)$; this is expected to be a reasonable
approximation even for $\epsilon=0.1$, which was considered in most simulations.
Guided by this analytical result, Fig. \ref{Wmin}(b) shows $Qt_{min}$ as a function 
of $l/\left(a\epsilon^3\right)$, which reasonably fits a straight line.
The slope of that fit is $0.16$, which is larger than, but of the same order of the theoretical slope $1/16$.

\subsection{Possible application to 2D materials growth}
\label{theoretical}

Here we discuss the application of the growth in the high coverage regime to produce 2D materials with small GB roughness.

Eqs. (\ref{Wminscaling}) and (\ref{WminL}) show that decreasing the growth temperature
(so that $\epsilon$ decreases) is advantageous for obtaining a smaller GB roughness.
However, Eq. (\ref{tminL}) shows that it is disadvantageous for increasing the time
to attain this minimal roughness because $Q$ and $\epsilon$ decrease.
Evidence on this disadvantage is provided by the comparison 
of Figs. \ref{roughness}(a) ($\epsilon=0.1$) and \ref{roughness}(c) ($\epsilon=0.05$) with
Fig. S.6 of the Supplementary Material ($\epsilon=0.01$).
The advantage involves a factor $\epsilon^{1/4}$ and the disadvantage involves
a factor $1/\left(Q\epsilon^3\right)$, so we generally expect that higher temperatures will be more
favorable for obtaining smoother GBs.
This implies that $\epsilon$ and $Q$ are large.
However, the condition $F/Q\gtrsim10$ is also necessary to
allow the initial uncorrelated growth of the separated grain edges (Sec. \ref{phasediagram}).

For the application to a particular 2D material, it is necessary that isolated grains nucleate
and grow to a configuration with smooth boundaries, which is the initial condition
considered here;
observe that the formation of this initial configuration is not addressed by our model.
In such configuration, a common situation is that the grains cover a fraction of the
substrate $\lesssim0.1$.
The inspection of microscopy images with visible and isolated islands (grains)
of 2D materials supports this estimate
\cite{dongAM2019,yaoACSNano2020,caiNanoRes2021,dongJPCL2021,hanNanoAdv2021,weiNanoRes2021}.
Instead, with coverages $\sim1$, coalescence of several islands may be observed.
Thus, if the separated grains have typical boundary length $L$, the gap width $l$ is expected to
be of a similar order of magnitude; for simplicity, the approximation $l\approx L$ is considered here.

Following the reasoning in the paragraphs above,
we also assume that:
edge lengths $L$ and initial gap widths $l$ are $\sim10\mu{\text m}$;
the lattice constant of the material is $a\approx 0.3$~nm
(which differs $\lesssim20\%$ from graphene and several metal dichalcogenides);
$\epsilon=0.1$, meaning kink detachment rate $10^2$ times smaller than
the attachment rate; the condition $F/Q\leq10$ is fulfilled.

From Eq. (\ref{Wmax}), the maximal roughness of the grain edges in these conditions is $\sim40$~nm.
However, Eq. (\ref{Wminscaling}) predicts that the minimal roughness attained by the GB
is $\sim2$~nm, corresponding to a decrease by a factor $\sim20$.
The time to attain the minimal roughness can be determined from Eq. (\ref{tminL}) in terms of the
attachment rate: $Qt_{min}\sim4\times{10}^6$.

For comparison, if near-equilibrium growth conditions are chosen and the GB relaxes
to its equilibrium configuration, the approach of Sec. S.III of the Supplementary Material
predicts the GB roughness
\begin{equation}
 W_{eq}=\sqrt{\frac{\epsilon aL}{6}} \left(\frac{1}{1-\epsilon}\right) .
\label{Weq}
\end{equation}
With the above parameters, we obtain $W_{eq}\sim 7$~nm, which is $3.5$ times larger than
the expected minimal roughness.
The time of crossover to the equilibrium roughness is obtained from the EW kinetics of the GB in
Sec. S.III of the Supplementary Material:
\begin{equation}
Qt_{eq}=\frac{\pi}{144}{\left(\frac{L}{a}\right)}^2 \frac{1}{\epsilon^2}
\left(\frac{1+\epsilon}{1-\epsilon}\right) .
\label{teq}
\end{equation}
It gives $Qt_{eq}\sim3\times{10}^9$, which exceeds by three orders of magnitude
the time to obtain the minimal roughness in the high coverage regime.

This comparison clearly shows that the advantages of growing 2d materials in the high coverage
regime can be extrapolated from our nanoscale simulations to more typical experimental conditions with
microscale grains.
The decrease of the GB roughness up to the minimum can be observed in a much shorter time than that
necessary for the GB to reach its equilibrium value, which is larger than that minimum.

Our model has the limitation that the mobile particle attachment is restricted to the
topmost point of each column $x$.
We noted that leads to unstable patterns that look
very different from those of experiments and of models for specific applications.
However, we believe that this limitation does not have significant effect on our
prediction of a minimal roughness for the high coverage regime.
If that condition for the attachment is relaxed, then the aggregation of mobile particles
to the sides of the rough edges may lead to overhang formation.
Such overhangs will encapsulate mobile particles, but these
particles are effectively static in the high coverage condition.
Thus, these particles will eventually attach at the same position and fill the overhang.
It is possible that this process creates some lateral correlations along the edges,
but such fluctuations, which are correlated at very small scales, can be suppressed approximately in
the same way as the totally uncorrelated fluctuations considered here.

\section{Conclusion}
\label{conclusion}

A model for the propagation of two grain edges of a 2d material
and for the relaxation of the grain boundary formed after the collision of those edges was proposed.
The model considers simple kinetic rules for
attachment and detachment of
particles from the edges and for particle diffusion while there is a gap between them.
This allowed to perform kinetic Monte Carlo (KMC) simulations for broad ranges of model
parameters and investigate their consequences
on the geometry of the grain edges.

Besides the well known near-equilibrium (attachment-limited) and unstable (diffusion-limited) growth regimes,
we observe the existence of a distinct third regime when the precursor flux is sufficiently 
high to fully cover the gap between the edges before they begin to grow.
In this high covered regime, an uncorrelated growth of the edges is observed until they collide.
Although this may produce an initial grain boundary with large roughness, we show that
this boundary relaxes to configurations with
a roughness smaller than those of the other regimes in a shorter time.
Numerical predictions of the minimal roughness and of the time to attain such configuration are
in good agreement with a theoretical approach for the grain boundary relaxation of the model.

To investigate the possible extension of the above results to real grains of 2d materials,
the simulation results (obtained at nanoscale) were extrapolated to edges and gaps
sizes $\sim10\mu{\text m}$.
It shows that the high coverage regime can produce grain boundaries with minimal roughness $\sim2$~nm,
which is $\sim3.5$ times smaller than the roughness obtained in a near-equilibrium, stationary state.
Moreover, the time to obtain that minimal roughness is three orders of magnitude smaller
than the expected time for the grain boundary to reach that stationary state.
Thus, if the control of the growth conditions of a 2d material permits to reach an initial high
coverage condition and a relaxation dynamics as proposed here, it is possible that a novel route for efficient growth of that material with very smooth grain boundaries can be developed.
Recent advances in microscopy methods allow accurate measurements of nanoscale fluctuations of GBs and may help such studies
\cite{annevelink2021,bokai2021}.

\section*{Acknowledgment}
FDAAR acknowledges support from the Brazilian agencies CNPq (305391/2018-6), FAPERJ 
(E-26/110.129/2013, E-26/202.881/2018), and CAPES (88887.310427/2018-00 - PrInt),
and from CNRS (France).
OPL acknowledges support from CAPES (88887.369967/2019-00).

\section*{Supplementary Material}

The Supplementary Material shows the possible temperature dependences of the model
parameters, additional images of the evolution of the grain boundaries,
additional plots of the roughness evolution, the Langevin model for grain boundary dynamics,
and the scaling of collision times in near-equilibrium and unstable regimes.

\vskip 1cm

\bibliographystyle{unsrt}

\newpage

\setcounter{section}{0}
\setcounter{figure}{0}
\setcounter{equation}{0}
\renewcommand\thesection{S.\Roman{section}}
\renewcommand\thefigure{S\arabic{figure}}
\renewcommand\theequation{S\arabic{equation}}

\section{Temperature dependence of model parameters}
\label{temperatureparameters}

The diffusion coefficient of particles that are not bonded to the grains can
be written as
\begin{equation}
D=a^2\nu\exp{\left( -E_s/k_BT\right)}
\label{defD}
\end{equation}
where $\nu$ is a frequency, $E_s$ is an activation energy, $k_B$ is the Boltzmann
constant and $T$ is the temperature.
In the deposition model of Clarke and Vvedensky \cite{cv}, $\nu$ is proportional
to the temperature.
However, several simulation works consider a constant value
$\nu\sim{10}^{12}{\text s}^{-1}$ and, under this assumption, obtain results
in good agreement with experiments on deposition of metals and semiconductors
\cite{etb}.
Typical values of $E_s$ for those materials are on the orders $0.1{\text{--}}1$~eV.

The attachment rate of particles at the edge borders may be written as
\begin{equation}
Q=Q_0\exp{\left( -E_a/k_BT\right)}
\label{defQ}
\end{equation}
where $Q_0$ is an attempt frequency and $E_a$ is the activation energy for that process.
These values are strongly dependent on the type of interaction between the 
atoms or molecules of the growing material.
In some cases, $Q$ may be considered as temperature independent, so $E_a=0$; this is the
case of several models of deposition of
metal and semiconductor films \cite{etb}.

The detachment probability can be written as
\begin{equation}
\epsilon = \exp{\left( -J/k_BT\right)} ,
\label{defepsilon}
\end{equation}
where $J$ denotes the bond energy with a NN.

\newpage 


\section{Morphological evolution with $Q/D={10}^{-3}$}
\label{morphologyRQ3}

\begin{figure}[!ht]
\center
\includegraphics[clip,width=\textwidth,angle=0]{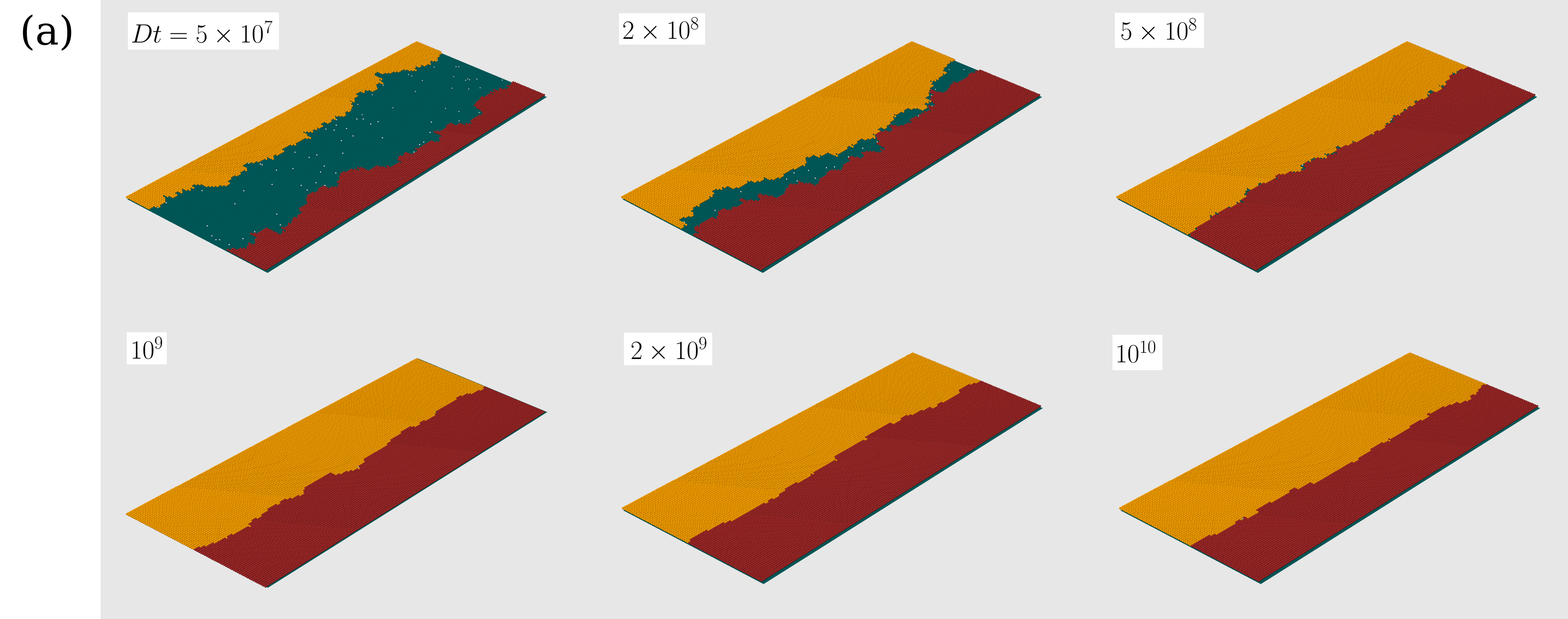}\\
\includegraphics[clip,width=\textwidth,angle=0]{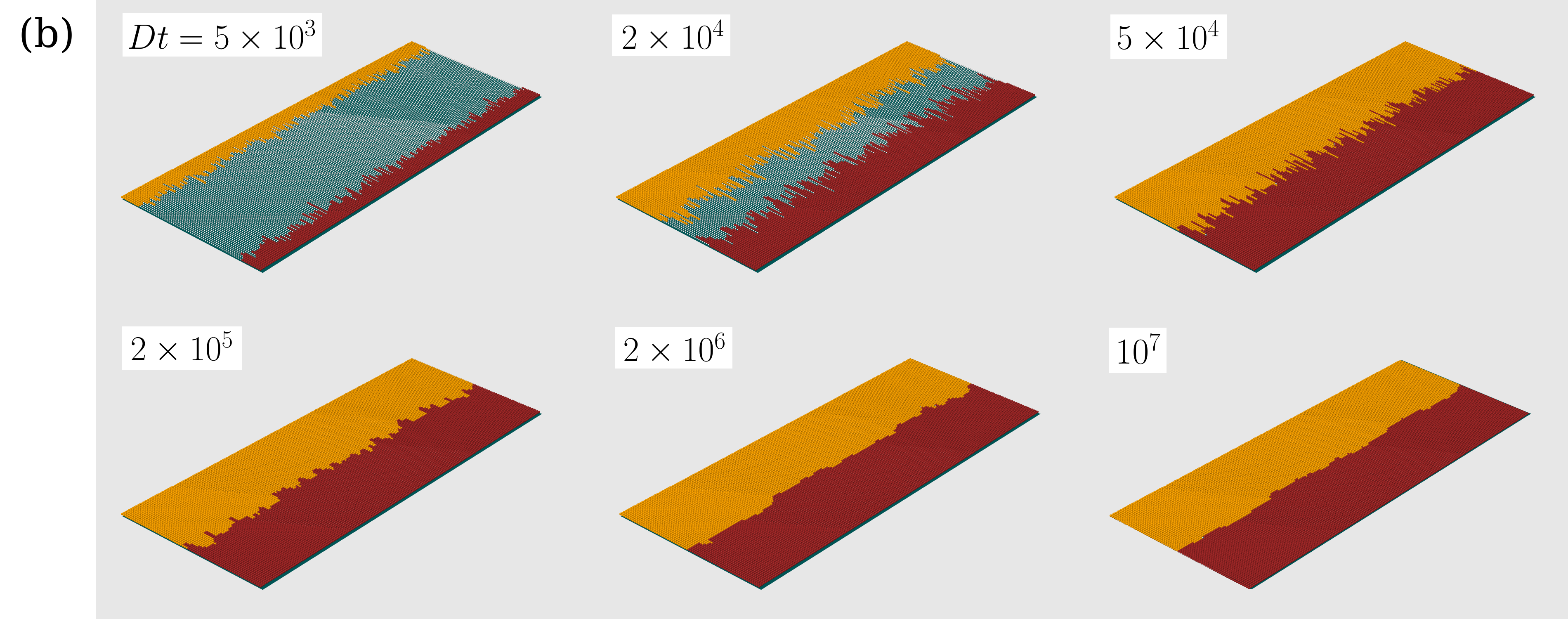}\\
\caption{
Snapshots of parts of the deposits grown with $Q/D={10}^{-3}$, $\epsilon=0.1$, $L=1024$, and $l=128$:
(a) $F/D={10}^{-8}$; (b) $F/D={10}^{-1}$.
}
\label{supimages}
\end{figure}

Fig. \ref{supimages}(a) shows snapshots of parts of the deposits grown with $F/D={10}^{-8}$, $Q/D={10}^{-3}$, $\epsilon=0.1$, $l=64a$, and
$L=1024a$.
In this parameter set, the value of $D$ is ${10}^2$ times larger than
that of Fig. 2a of the main text, with the other parameters kept at the
same values.
It is an example of edge growth in the attachment-limited regime.

Fig. \ref{supimages}(b) shows snapshots of the deposits grown with $F/D={10}^{-1}$,
$Q/D={10}^{-3}$, $\epsilon=0.1$, $l=64$, and
$L=1024$.
The value of $D$ is ${10}^2$ times larger than
that of Fig. 2c of the main text, with the other parameters kept at the
same values.
It is an example of edge growth in the high coverage regime.

\newpage

\section{Langevin model for GB dynamics}
\label{Appendix_Relaxation_after_collision}

In this section, we propose a Langevin model
for the dynamics of the GB, when all sites
are occupied by atoms from the two grains.
The dynamics is well described by two types of rates.
The first one accounts for the rate $R_+$ at which
an atom detaches from the upper crystal and reattaches 
to the lower crystal. Since the detachment rate is
$\epsilon^{4-n}$ where $n$ is the number of in-plane
nearest neighbors with the lower crystal, and 
the attachment rate is $Q/2$, we have
\begin{align}
    R_+(n)&=\frac{Q}{2}\epsilon^{4-n}.
\end{align}
Similarly, the rate of detachment from the lower crystal
and attachment to the upper crystal is
\begin{align}
    R_-(n)&=\frac{Q}{2}\epsilon^{n}.
\end{align}

\subsection{Derivation from Einstein's relation} 
Since atomic steps are 1D systems with short range interactions,
they are always in a high-temperature phase, i.e., they are always rough
and cannot undergo a low-temperature facetting transition. 
Hence, there is always a finite concentration of kinks.
At low temperatures, their concentration is small and is given by the Boltzmann weight
of non-interacting excitations of energy $E_k$. This leads to $p_k=2\exp[-E_k/k_BT]$,
where $E_k=J$ is the kink energy, and the factor $2$ comes from the
existence of two types of kinks (up and down kinks when going along $+x$).

Using a model with kinks of arbitrary heights (but imposing the SOS constraint that neglects
the possibility of overhangs as in KMC) the kink density reads~\cite{saitobook}
\footnote{We use the relation $\langle |n| \rangle=\partial_{J_x}\beta_0$
in Eq.10.33 of Ref. \protect\cite{saitobook}, and then $J_x\rightarrow J$.}
\begin{align}
p_k= \langle |n| \rangle  =\frac{2 \mathrm{e}^{-J/k_BT}}{1- \mathrm{e}^{-2J/k_BT}} = \frac{2\epsilon}{1-\epsilon^2}\,.
\end{align}
In addition, the hopping rate of a kink to the left or to the right is $R_+(2)=R_-(2)=R(2)$,
so that the diffusion constant of a kink along the $x$ axis parallel to the (10) direction is 
\begin{align}
    D_k=a_{\parallel}^2R(2)=a_{\parallel}^2\frac{Q}{2}\epsilon^2\,,
    \label{eq:Dk_diffusion}
\end{align}
where we have defined the lattice parameter $a_{\parallel}$
parallel to the GB ($a_{\parallel}=a$ in our simulations). From Einstein's relation,
the mobility of a kink is
\begin{align}
    M_k=\frac{D_k}{k_BT}=\frac{a_{\parallel}^2R(2)}{k_BT}\,.
\end{align}
In the presence of a kink chemical potential $\mu_k$,
the velocity of a kink is assumed to obey a linear kinetic
law
\begin{align}
    v_k=-M_k\frac{ \mu_k}{a_{\parallel}}\, .
\end{align}

We now wish to make link with macroscopic quantities.
The macroscopic position of the GB is $h_{GB}(x,t)$.
Its macroscopic velocity
~\footnote{A simple derivation of this equation follows.
We first define the number of atoms under the interface
\begin{align}
    {\cal N}=\int \mathrm d x \;\frac{h_{GB}(x,t)}{\Omega}.
\end{align}
where $\Omega=a_{\parallel}a_{\perp}$.
As a consequence
\begin{align}
    \partial_t{\cal N}=\int \mathrm d x \; \frac{\partial_t h_{GB}(x,t)}{a_{\parallel}a_{\perp}}.
\end{align}
The change in the number of atoms can also be written in terms of 
the frequency $v_k/a_{\parallel}$ of addition or removal of atoms at kinks:
\begin{align}
    \partial_t{\cal N}=\sum_{\mathrm{kinks}}\frac{v_k}{a_{\parallel}}.
\end{align}
Then, we switch to a summation on the index $i$ of sites along the $x$ axis
\begin{align}
    \partial_t{\cal N}=\sum_{i}p_k\frac{v_k}{a_{\parallel}}.
\end{align}
Finally, we take the continuum limit where $\sum_{i}a_{\parallel}\rightarrow \int \mathrm d x$:
\begin{align}
    \partial_t{\cal N}= \int \mathrm d x\; p_k\frac{\langle v_k\rangle }{a_{\parallel}^2}.
\end{align}
where $\langle v_k\rangle$ is the average of $v_k$ 
over some mesoscopic lengthscale.
Indentification of the two writings of $\partial_t{\cal N}$
leads to Eq.(\ref{eq:maco_edge_veloc_kinks0}).
}
then reads~\cite{Ghez1988,Caflisch1999}
\begin{align}
    \partial_th_{GB}=a_{\perp}\langle v_k \rangle\frac{p_k }{a_{\parallel}} ,
    \label{eq:maco_edge_veloc_kinks0}
\end{align}
where $a_{\perp}$ is the lattice constant in the direction perperdicular to the GB.
Note that Eq.(\ref{eq:maco_edge_veloc_kinks0}) was  derived in Refs.\cite{Ghez1988,Caflisch1999}
under the assumption that all kinks 
have the same velocity during growth. However, this relation is also valid 
 at equilibrium
where the motion of kinks is of the diffusion type with zero average.
The relation between these two approaches is an example 
of the usual Einstein close-to equilibrium link
between mobility for macroscopic motion under under fixed
external force and diffusion at equilibrium.

Since $M_k$ does not vary from one kink to the other,
we have~\cite{Misbah2010}
\begin{align}
    \langle v_k\rangle =-M_k\frac{ \langle\mu_k\rangle}{a_{\parallel}}
\end{align}
and, since the chemical potential represents the 
free energy gain upon the addition of a single particle,
we expect that the microscopic kink chemical potential and the macroscopic
chemical potential to be equal:
\begin{align}
     \langle\mu_k\rangle = \mu .
\end{align}
In addition, we use the standard macroscopic law
\begin{align}
    \mu= \frac{\delta{\cal F}/\delta h_{GB}}{\delta{\cal N}/\delta h_{GB}}=\Omega \tilde\gamma_{GB}\kappa
    \approx - \Omega \tilde\gamma_{GB}\partial_{xx}h_{GB}\, ,
\end{align}
where: ${\cal F}$ and ${\cal N}$ are respectively the edge free energy 
and the number of atoms in the lower crystal;
$\Omega=a_{\parallel}a_{\perp}$ is the atomic area;
$\tilde{\gamma}_{GB}$ is the grain boundary stiffness, obtained by the substitution $J\rightarrow 2J$
into the usual expression of the stiffness of a lattice model with bond energy $J$
(see, e.g. Ref. \protect\cite{saitobook}),
leading to:
\begin{align}
    \tilde{\gamma}_{GB} = \frac{k_B T a_{\parallel}}{2a_{\perp}^2} \big( \varepsilon^{-1/2} - \varepsilon^{1/2} \big)^2.
    \label{eq:stiffness_GB}
\end{align}

Finally, we obtain
\begin{align}
    \partial_th_{GB}=- \frac{a_{\perp}}{a_{\parallel}} \frac{M_k \mu}{a_{\parallel}} p_k
    =   \frac{R(2) p_k}{k_BT}  a_{\perp} \Omega \tilde\gamma_{GB}\partial_{xx}h_{GB} \, .
    \label{eq:maco_edge_veloc_kinks}
\end{align}
Adding a Langevin force $\eta_{GB}$, we obtain an EW equation
\begin{align}
    \partial_th_{GB} =
     R(2)p_k a_{\perp} \Gamma_{GB}\partial_{xx}h_{GB} +\eta_{GB} ,
    \label{eq:Langevin_edge_veloc_kinks}
\end{align}
with $\Gamma_{GB}=\Omega \tilde\gamma_{GB}/\left( k_BT\right)$.
Using again the fluctuation dissipation theorem to determine the amplitude of the equilibrium Langevin force $\eta(x,t)$ (following the same lines as in Refs. \protect\cite{PierreLouis1998,Misbah2010}), we obtain
the roughness of the GB as
\begin{align}
    W_{GB}^2&=\frac{2}{L^2}\sum_{n>0}\langle |h_q(0)|^2\rangle {\rm e}^{-2R(2)p_ka_{\perp}\Gamma_{GB} q^2 t}
    \nonumber \\
    &+\frac{L\Omega}{2\pi^2\Gamma_{GB}}\sum_{n>0}\frac{1}{n^2}
    \left(1-{\rm e}^{-2R(2)p_ka_{\perp}\Gamma_{GB} q^2 t}\right) ,
    \label{eq:roughness_GB_Langevin}
\end{align}
where $h_q\left(t\right)$ denotes the Fourier mode with $q=2n\pi/L$.
At long times, we recover the well known expression of the equilibrium (stationary) roughness~\cite{saitobook,Misbah2010}
\begin{align}
W_{eq}^2=\frac{\Omega L}{12 \Gamma_{GB}}\, ,
\end{align}
and using Eq.(\ref{eq:stiffness_GB}) we obtain
\begin{align}
W_{eq}^2=\frac{ L a_\perp^2}{6a_\parallel}\,\frac{\epsilon}{(1-\epsilon)^2}\, .
\label{eq:Weq_GB}
\end{align}

At short times, we find
\begin{align}
    W_{GB}^2&=\frac{2}{L^2}\sum_{n>0}\langle |h_q(0)|^2\rangle {\rm e}^{-2R(2)p_ka_{\perp}\Gamma_{GB} q^2 t}
    \nonumber \\
    &+\Omega\left(\frac{2R(2)p_k a_{\perp}}{\pi \Gamma_{GB}} t
    \right)^{1/2}\, .
    \label{eq:roughness_GB_Langevin_EW_and_relax}
\end{align}
The second term of this equation is the well known equilibrium
EW scaling. It is e.g. in agreement with Eqs.(2.63,2.65) of Ref. \protect\cite{Misbah2010} 
with the substitution 
$(\nu_++\nu_-)\Omega c_{eq}^0\rightarrow R(2)p_ka_{\perp}$.

We substitute the expressions of $p_k$ and $\tilde\gamma_{GB}$ as a function of $\epsilon$ and obtain
\begin{align}
    W_{GB}^2&=\frac{2}{L^2}\sum_{n>0}\langle |h_q(0)|^2\rangle {\rm e}^{-2R(2)p_ka_{\perp}\Gamma_{GB} q^2 t}
    \nonumber \\
    &+ \Omega \left(\frac{4\epsilon^4 a_{\perp}^2}{\pi(1+\epsilon)(1-\epsilon)^3a_{\parallel}^2}  Qt
    \right)^{1/2}\, .
    \label{eq:roughness_GB_Langevin_epsilon}
\end{align}

\subsection{Minimal roughness for fast growth with coverage $\theta=1$}

The roughness in Eq. (\ref{eq:roughness_GB_Langevin_EW_and_relax}) 
can exhibit a minimum. We re-write this equation as
\begin{align}
    W_{GB}^2&=\frac{2}{L^2}\sum_{n>0}\langle |h_q(0)|^2\rangle {\rm e}^{-bn^2 t}
    +\frac{L\Omega}{2\pi\Gamma_{GB}}\left(\frac{bt}{\pi} 
    \right)^{1/2}\, ,
    \label{eq:roughness_GB_Langevin_b}
\end{align}
where 
\begin{align}
    b&= 2R(2)p_k a_{\perp}\Gamma_{GB} \left(\frac{2\pi}{L}\right)^2\, .
\end{align}
The evolution of $W_{GB}^2$ depends on the details of the initial condition
which dictates the power spectrum $\langle |h_q(0)|^2\rangle$ just after collision. 

In the limit of fast growth with coverage $1$, we assume that 
the growth regime before collision corresponds
to RD, with forward atomic moves
of the two interfaces appearing with a frequency $Q$.
This gives a collision time $t_0=l/(2a_{\perp}Q)$
and a roughness before collision $W_{1bc}^2=a_{\perp}^2Qt_0=a_{\perp}l/2$. 
As discussed in Ref. \protect\cite{reispierrelouis2018}, the collision
reduces the square roughness by a factor $2$, so that after the collision
$W^2_{1ac}=W_{1bc}^2/2=a_{\perp}l/4$. 
In the RD regime, the roughness is spatially uncorrelated, so
\begin{align}
    \langle h_{ac}(x) h_{ac}(x') \rangle 
    = W_{1ac}^2 a_{\parallel}\;\delta(x-x')
\end{align}
and, in Fourier space, 
\begin{align}
    \langle h_{ac,q} h_{ac,q'} \rangle 
    = a_{\parallel}W_{1ac}^2 \;2\pi\delta(q-q')
    = a_{\parallel}W_{1ac}^2 \;L\delta_{n+n'}\, .
\end{align}
As a consequence, $\langle |h_q(0)|^2\rangle=a_{\parallel}W_{1ac}^2 \;L={\Omega}lL/4$ in Eq.(\ref{eq:roughness_GB_Langevin_b}),
leading to
\begin{align}
    W_{GB}^2&=\frac{2 a_{\parallel}W_{1ac}^2}{L}\sum_{n>0} {\rm e}^{-bn^2 t}
    +\frac{L \Omega}{2\pi\Gamma_{GB}}\left(\frac{bt}{\pi} \right)^{1/2}
    \nonumber \\
    &=\frac{2 a_{\parallel}W_{1ac}^2}{L}\Theta_3({\rm e}^{-b t})
    +\frac{L \Omega}{2\pi\Gamma_{GB}}\left(\frac{bt}{\pi} \right)^{1/2}\, ,
    \label{eq:roughness_GB_Langevin_cov1_b}
\end{align}
where $\Theta_3$ is the Elliptic Theta function.
For $t\gg1/b$, we have ${\rm e}^{-b t}\ll 1$, $\Theta_3({\rm e}^{-b t})\approx {\rm e}^{-b t}$, and
\begin{align}
    W_{GB}^2&=\frac{2 a_{\parallel}W_{1ac}^2}{L} {\rm e}^{-b t}
    +\frac{L \Omega}{2\pi\Gamma_{GB}}\left(\frac{bt}{\pi} \right)^{1/2}\, ,
    \label{eq:roughness_GB_Langevin_cov1_b_largetimes}
\end{align}
i.e., the sum is dominated by the slowest decaying mode $n=1$.
This expression does not necessarily lead 
to a minimum; the condition of existence of a minimum
of the function ${\rm e}^{-x}+\alpha x^{1/2}$
is that $\alpha<2^{1/2}{\rm e}^{-1/2}$.
Using $x=bt$, we obtain the condition for the presence of a minimum
\begin{align}
    \alpha&=\frac{L^2 \Omega}{4\pi^{3/2}W_{1ac}^2\Gamma_{GB} a_{\parallel}}<2^{1/2}{\rm e}^{-1/2} \, ,
\end{align}
which is re-written using the relation $W^2_{1ac}=a_{\perp}l/4$ as
\begin{align}
    L  < L_{\mathrm{sup}} 
    &= \left(\frac{2\pi^3}{\mathrm e}\right)^{1/4}(l\Gamma_{GB})^{1/2}
    \label{eq:Lsup}
    \\
    & = \left(\frac{\pi^3}{2\mathrm e}\right)^{1/4}\left( \frac{l }{a_{\perp}} \frac{(1-\epsilon)^2}{\epsilon}\right)^{1/2}a_{\parallel}\, .
    \nonumber
\end{align}

In the opposite regime $bt\ll 1$, when the time is 
to short to allow for the relaxation of the modes with wavelength close to the system size $L$,
but long enough for short wavelength
modes to have decayed (so that the sum on $n$ can be taken up to infinity), one can take
the continuum limit in $n$:
\begin{align}
    \sum_{n>0} {\rm e}^{-bn^2 t}\approx\int_0^\infty \mathrm dn\; {\rm e}^{-bn^2 t}
    =\frac{1}{2}\left(\frac{\pi}{bt}\right)^{1/2}
\end{align}
up to an additive constant related to the accuracy of the continuum limit around $n=0$. 
This leads to
\begin{align}
    W_{GB}^2&=\frac{ a_{\parallel}W^2_{1ac}}{L}\left(\frac{\pi}{bt}\right)^{1/2}
    +\frac{L\Omega}{2\pi\Gamma_{GB}}\left(\frac{bt}{\pi} \right)^{1/2}
    \nonumber \\
    &=\frac{ a_{\parallel}W^2_{1ac}}{\tau^{1/2}}+\frac{\Omega\tau^{1/2}}{2\pi\Gamma_{GB}}\, ,
    \label{eq:roughness_GB_Langevin_cov1_b_shorttimes}
\end{align}
where $\tau=L^2bt/\pi$.

A minimum then is found for $\tau_{min}=2\pi\Gamma_{GB} W_{1ac}^2 a_{\parallel}/\Omega$, leading to
\begin{align}
    t_{min}&=\frac{W_{1ac}^2}{4R(2)p_k a_{\perp}^2}
  = \frac{l}{16 R(2)p_k a_{\perp}}\, .
    \label{eq:roughness_GB_Langevin_cov1_b_shorttimes_min}
\end{align}
Interestingly, $t_{min}$ does not depend on the stiffness $\tilde{\gamma}_{GB}$.
This minimum is obtained for $bt_{min}=2\pi^2 a_{\parallel}\Gamma_{GB} W_{1ac}^2/(L^2 \Omega)=\pi^2\Gamma_{GB} l/\left( 2L^2\right)\ll1$,
leading to the condition 
\begin{align}
L>L_{\mathrm{inf}}=\pi (\Gamma_{GB} l/2)^{1/2}.
\label{eq:Linf}
\end{align}
Note the close similarity between the expressions
of $L_{\mathrm{inf}}$ in Eq.(\ref{eq:Linf}) and $L_{\mathrm{sup}}$ in Eq.(\ref{eq:Lsup}),
which are identical up to their prefactors [which are numerically
similar: $\pi\approx 3.14$ and $(8\pi^3/\mathrm{e})^{1/4}\approx 3.09$].
The value of the roughness at the minimum is
\begin{align}
    W_{min}^2&=\frac{2W_{1ac}( a_{\parallel}\Omega)^{1/2}}{(2\pi\Gamma_{GB})^{1/2}}
    =\Omega\left(\frac{l}{2\pi\Gamma_{GB}}\right)^{1/2} \\
    &=\frac{\Omega}{1-\epsilon}\left(\frac{\epsilon la_{\perp}}{\pi a_{\parallel}^2}\right)^{1/2}=\left(\frac{\epsilon a_{\perp}^3}{\pi(1-\epsilon)^2}\right)^{1/2}l^{1/2}\, .
\end{align}
As a remark, $W_{min}^2$ does not depend on the kinetic coefficient $R(2)$.

Finally, the total physical time for the minimum is
\begin{align}
    t_{M}&=t_0+t_{min}=\frac{l}{2a_{\perp}Q}\left(1+
    \frac{1}{4\epsilon^2 p_k}
    \right)
\nonumber \\
    &=\frac{l}{2a_{\perp}Q}\left(1+
    \frac{(1-\epsilon^2)}{8\epsilon^3}\right)\, .
\end{align}
Again, note that $t_M$ does not depend on the stiffness.
We also see that $t_M\gg t_0$ for $\epsilon\ll 1$.

As a summary, for large $L$, we expect a minimum
at a time $t_M$ that does not depend on the energetic properties of the interface,
with a roughness $W_{min}^2$ that does not depend on the kinetics of the interface.



\subsection{Crossover between EW regime and equilibrium}
\label{crossover}

Here, we wish to evaluate the crossover time between the EW growth of the GB roughness
and the equilibrium regime that occurs at long times. From Eq.(\ref{eq:roughness_GB_Langevin_b}), the 
growth of the GB roughness in the EW regime reads
\begin{align}
W_{GB,EW}^2 =  \Omega \left(\frac{2R(2)p_ka_\perp t_{eq}}{\pi\Gamma_{GB}}\right)^{1/2}
\frac{\Omega L}{12\Gamma_{GB}}
\end{align}
From the condition $W_{GB,EW}^2=W_{eq}^2$ at $t=t_{eq}$, and using
Eq.(\ref{eq:Weq_GB}),
we obtain
\begin{align}
Qt_{eq}=\frac{\pi}{144}{\left(\frac{L}{a}\right)}^2 \frac{1}{\epsilon^2}\,
\left(\frac{1+\epsilon}{1-\epsilon}\right) .
\label{eq:crossover time}
\end{align}

An alternative definition of the crossover time is 
that the decay of the slowest mode $n=1$ 
in Eq.(\ref{eq:roughness_GB_Langevin_b}) takes place, leading to $bt_{eq}^*=1$.
This conditions can be rewritten as $t_{eq}^*=(36/\pi^3)t_{eq}\approx 1.16 t_{eq}$.
As a consequence, this other definition of the crossover time
is very similar to the one defined above and used in the main text.

\newpage

\section{Roughness evolution for initial gap width $l=32$}
\label{gap32}

\begin{figure}[!ht]
\center
\includegraphics[clip,width=\textwidth,angle=0]{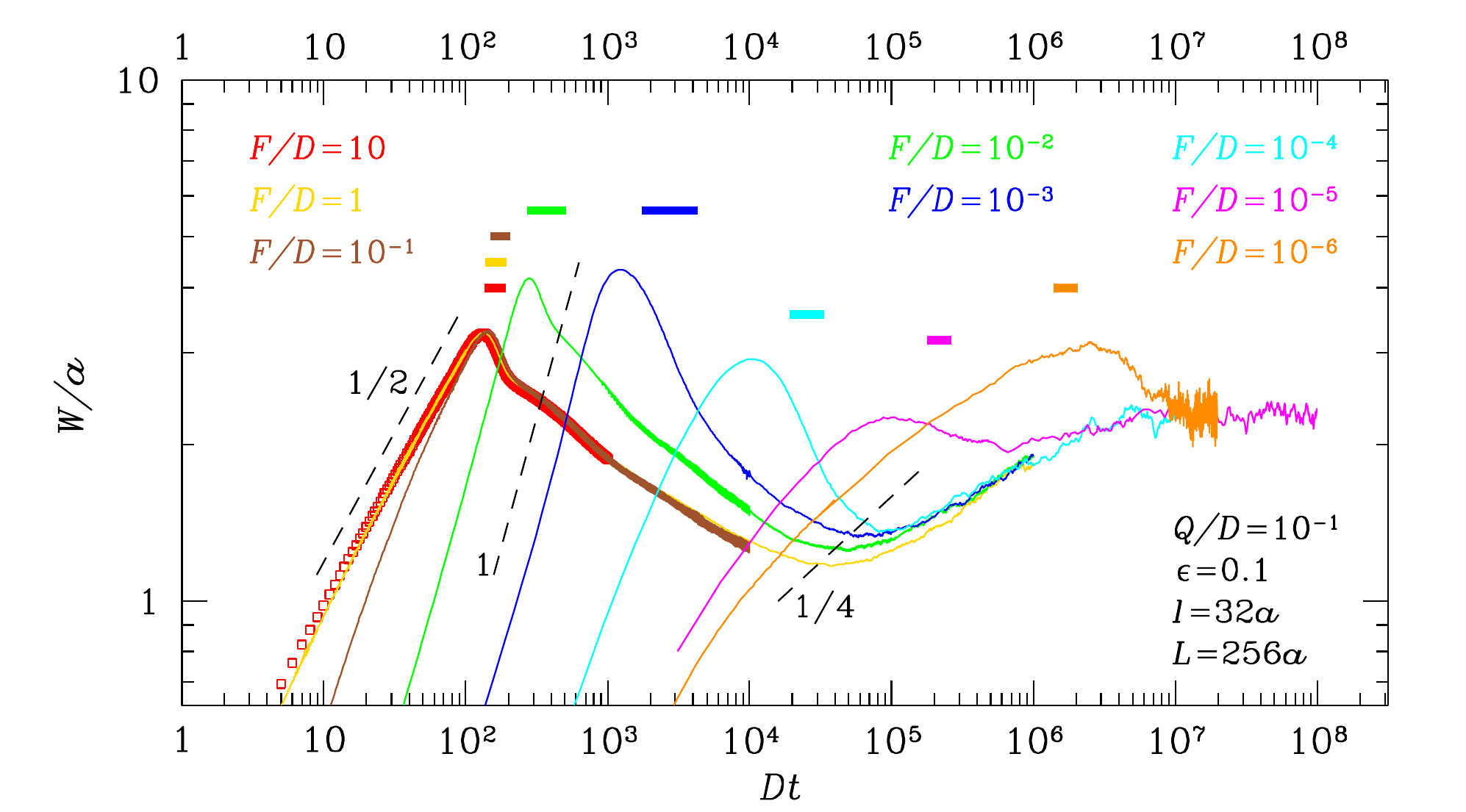}
\caption{
Roughness evolution for initial gap width $l=32$.
Dashed lines have the indicated slopes and the 
horizontal solid bars indicate the range of collision times (average plus and minus one standard deviation).
}
\label{roughness32}
\end{figure}

Fig. \ref{roughness32} shows the roughness evolution for the same deposition parameters
of Fig. 4a of the main text [$Q/D={10}^{-1}$, $\epsilon=0.1$, and several flux rates $F$],
but gap width $l=32$ and edge length $L=256$.

The smaller values of the roughness before and after the formation
of the GB in Fig. \ref{roughness32}, compared with Fig. 4a of the main text,
are consequences of the decrease of the gap width $l$.

The change in the edge length $L$ neither affects the initial roughening nor the
relaxation after the GB formation because there are not effects of
the finite lateral size in those regimes.
As explained in Sec. SI.III, this length affects only the stationary roughness, which increases as $L^{1/2}$.

\newpage


\section{Collision time in the high coverage regime}
\label{collisionhighcoverage}

\begin{figure}[!ht]
\center
\includegraphics[clip,width=0.5\textwidth,angle=0]{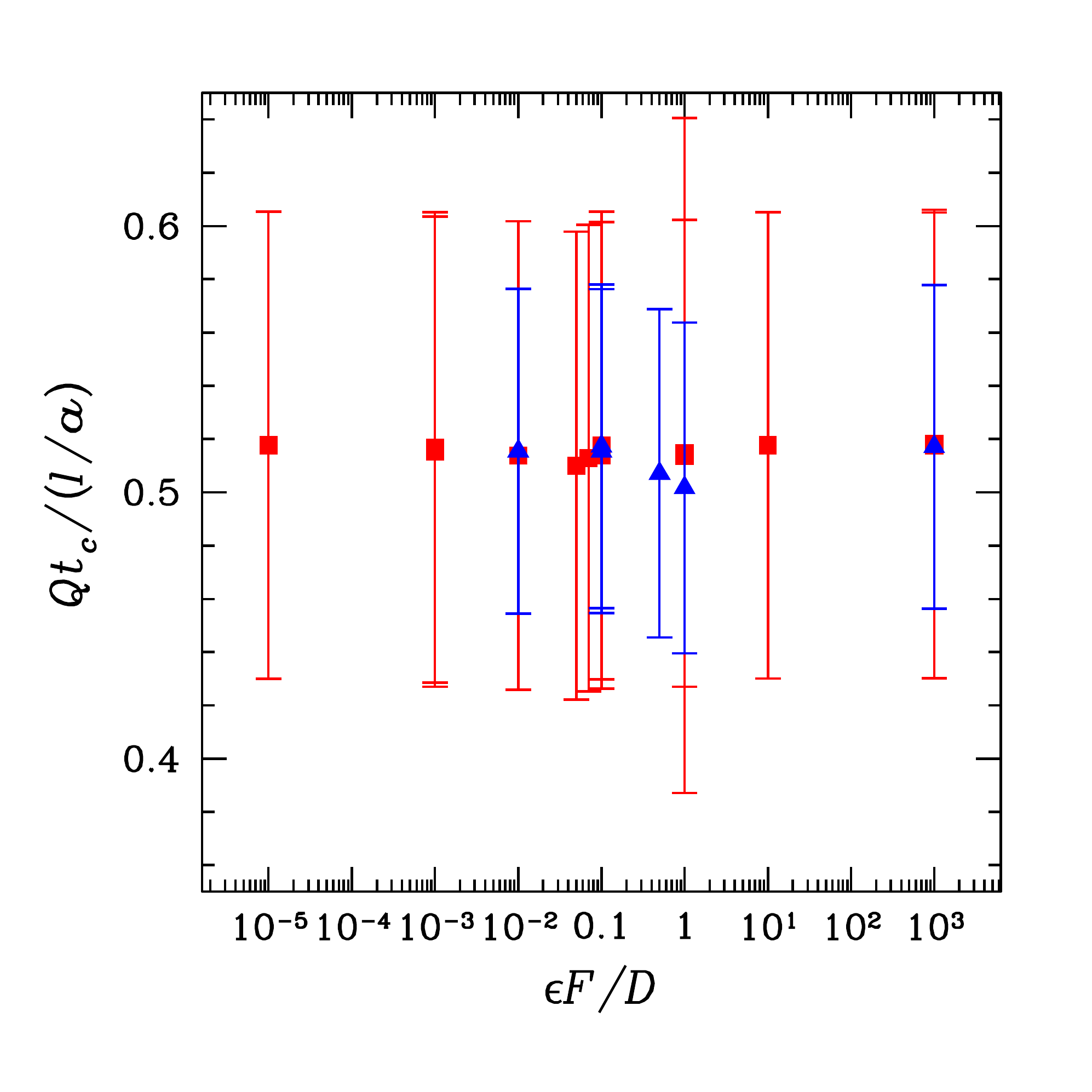}
\caption{
Scaled collision time as a function of the flux-diffusion ratio $F/D$.
Different colors distinguish the different gap widths $l$.
}
\label{tcol}
\end{figure}

Fig. \ref{tcol} shows the collision times scaled by the attachment rate $Q$ and by the
dimensionless gap width $l/a$, as a function of $\epsilon F/D$.
The data were obtained for different values of the detachment rate $0.01\leq\epsilon\leq0.1$
and in different edge lengths $L$.
This scaled plot shows that
\begin{equation}
t_c\sim 0.5 \frac{l/a}{Q} ,
\label{tcscaling}
\end{equation}
independently of the other 
model parameters.

\newpage


\section{Roughness evolution for $Q/D={10}^{-3}$}
\label{RQ3}

Fig. \ref{roughness32} shows the roughness evolution for $Q/D={10}^{-3}$ and several flux rates $F$,
with the other parameters equal to those of
Fig. 2a of the main text.

\begin{figure}[!ht]
\center
\includegraphics[clip,width=\textwidth,angle=0]{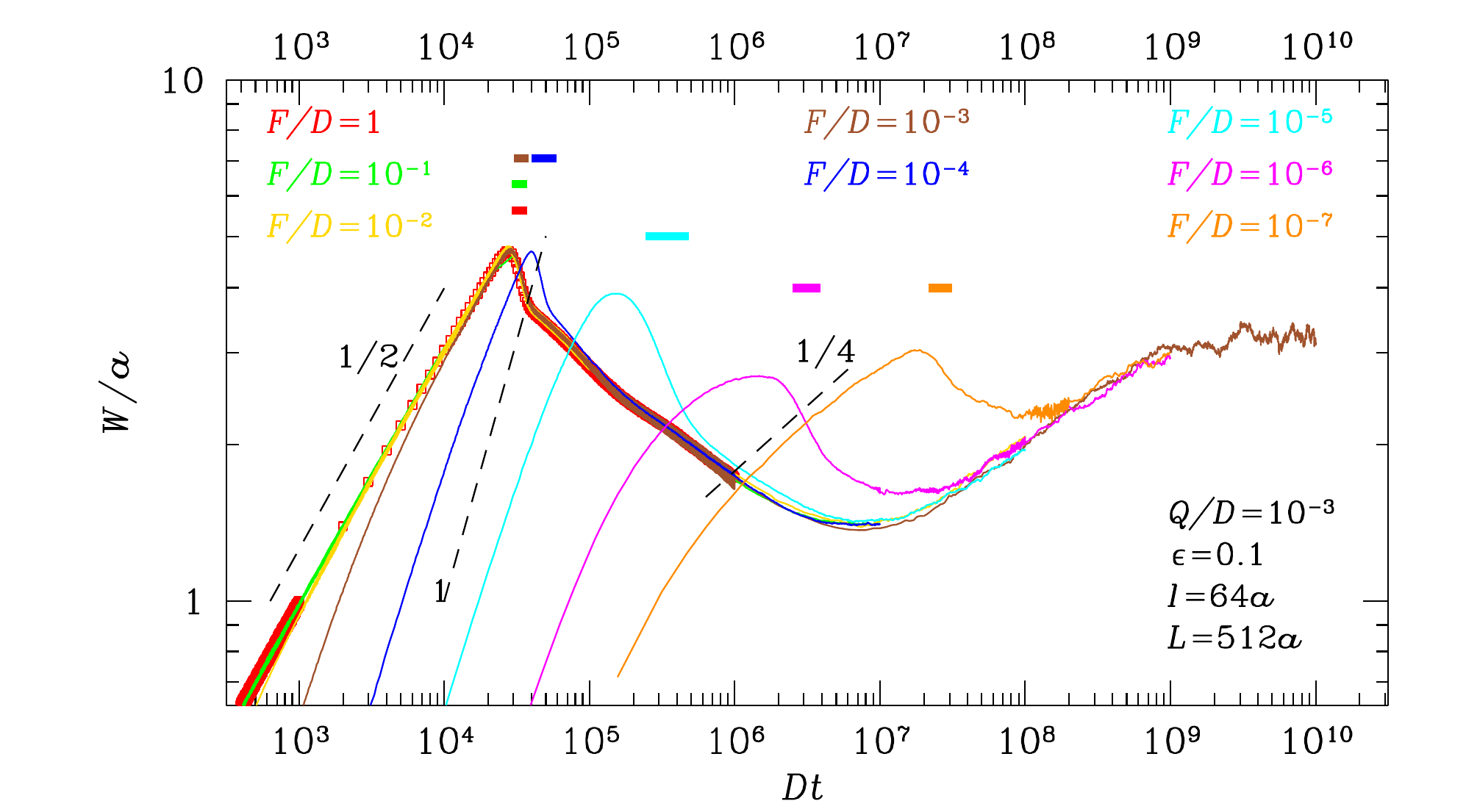}
\caption{
Roughness evolution for several values of
the flux.
Dashed lines have the indicated slopes and horizontal solid bars indicate the collision times (average plus and minus one standard deviation).
}
\label{wpd}
\end{figure}

This is a case of large diffusion coefficient of particles in the gap in comparison with the attachment rate.
It suppresses the unstable regime (diffusion-limited), as revealed by the
absence of a slope larger than $1$ in those plots.


\newpage

\section{Roughness evolution for small detachment rate ($\epsilon=0.01$)}
\label{eps001}

\begin{figure}[!ht]
\center
\includegraphics[clip,width=0.5\textwidth,angle=0]{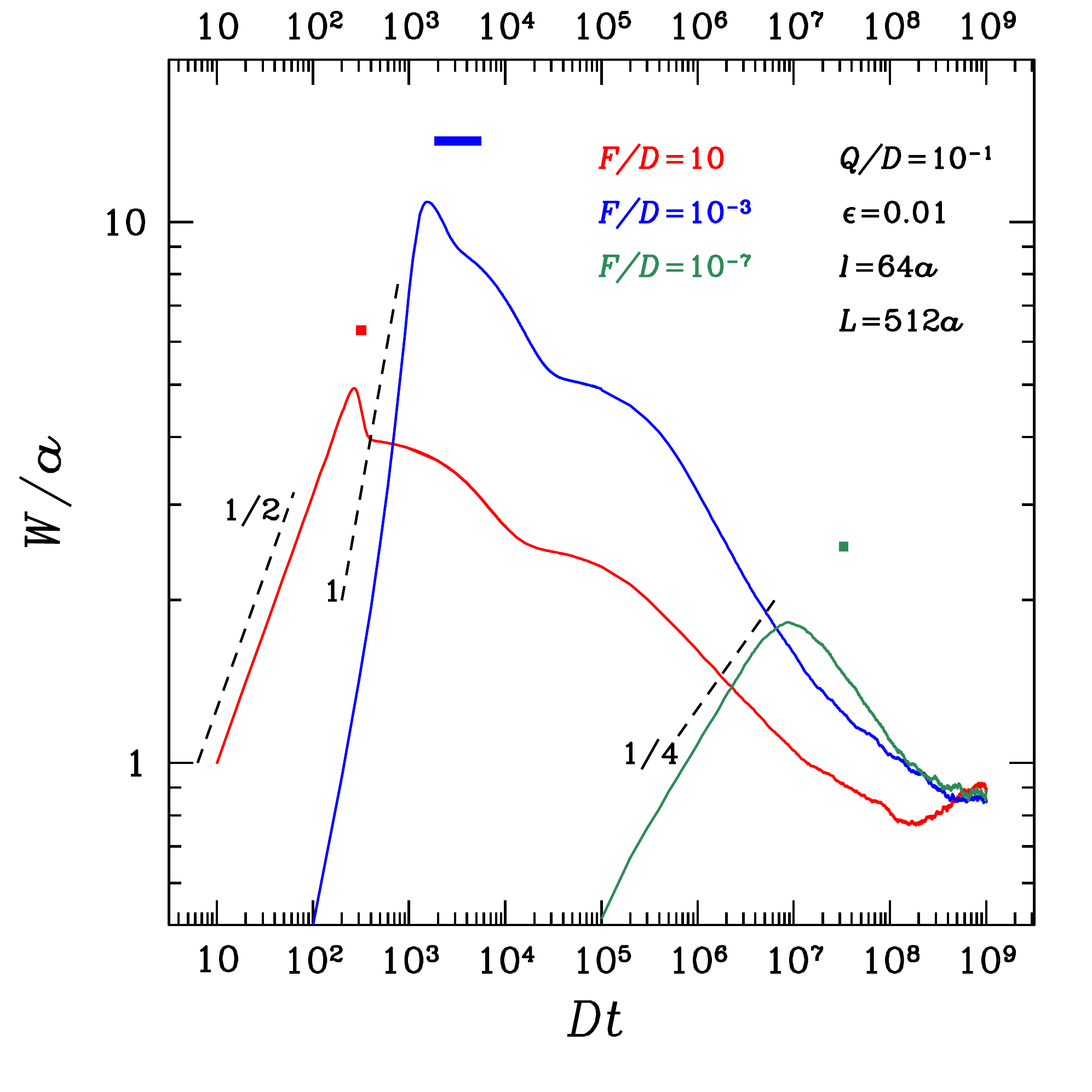}
\caption{
Roughness evolution with small detachment rate, for several values of the flux.
Dashed lines have the indicated slopes.
}
\label{wpe}
\end{figure}

Fig. \ref{wpe} shows the roughness evolution for three flux rates $F$,
with the same parameters of Fig. 2a of the main text except $\epsilon=0.01$.

\end{document}